\documentclass[twocolumn,twocolappendix]{aastex6}
\shorttitle{Symmetric Achromatic Variability}
\shortauthors{Vedantham et al.}
\begin{document}
\title{Symmetric Achromatic Variability in Active Galaxies -- A Powerful New Gravitational Lensing Probe?}
\author{H. K. Vedantham\altaffilmark{1}, 
A. C. S. Readhead\altaffilmark{1}, 
T. Hovatta\altaffilmark{2,3,4},
T. J. Pearson\altaffilmark{1}, 
R. D. Blandford\altaffilmark{5}, 
M. A. Gurwell\altaffilmark{6}, 
A. L\"ahteenm\"aki\altaffilmark{2}, 
W. Max-Moerbeck\altaffilmark{7}, 
V. Pavlidou\altaffilmark{8}, 
V. Ravi\altaffilmark{1}, 
R. A. Reeves\altaffilmark{9}, 
J. L. Richards\altaffilmark{1}, 
M. Tornikoski\altaffilmark{2}, 
J. A. Zensus\altaffilmark{7}}
\altaffiltext{1}{Owens Valley Radio Observatory, California Institute of Technology,  Pasadena, CA 91125, USA} % Vedantham, Readhead, Pearson, Richards
\altaffiltext{2}{Aalto University Mets\"ahovi Radio Observatory,  Mets\"ahovintie 114, 02540 Kylm\"al\"a, Finland} % Hovatta, Lahtneenmaki
\altaffiltext{3}{Aalto University Department of Radio Science and Engineering, Finland} % Hovatta, Lahtneenmaki
\altaffiltext{4}{Tuorla Observatory, Department of Physics and Astronomy,  University of Turku, Finland} % Hovatta
\altaffiltext{5}{Kavli Institute for Particle Astrophysics and Cosmology, Department of Physics, and  SLAC National Accelerator Laboratory, Stanford University, Stanford, CA 94305, USA} % Blandford
\altaffiltext{6}{Harvard-Smithsonian Center for Astrophysics, Cambridge, MA 02138, USA} % Gurwell
\altaffiltext{7}{Max-Planck-Institut f\"ur Radioastronomie, Auf dem H\"ugel 69, D-53121 Bonn, Germany} % Max-Moerbeck
\altaffiltext{8}{Department of Physics and Institute of Theoretical and Computational Physics, University of Crete, 71003 Heraklion, Greece, and Foundation for Research and Technology -- Hellas, IESL, 7110 Heraklion, Greece} % Pavlidou
\altaffiltext{9}{CePIA, Astronomy Department, Universidad de Concepci\'on,  Casilla 160-C, Concepci\'on, Chile} % Reeves
\begin{abstract}
We report the discovery of a rare new form of long-term radio variability in the light-curves of active galaxies (AG)  --- Symmetric Achromatic Variability (SAV) --- a pair of opposed and strongly skewed peaks in the radio flux density observed over a broad frequency range. We propose that SAV arises through gravitational milli-lensing when relativistically moving features in AG jets move through gravitational lensing caustics created by  $10^3-10^6 \;{\rm M}_{\odot}$ subhalo condensates or black holes located within intervening galaxies. The lower end of this mass range has  been inaccessible with previous gravitational lensing techniques.  This new interpretation of some  AG variability can easily be tested and if it passes these tests, will enable a new and powerful probe of cosmological matter distribution on these intermediate mass scales, as well as provide, for the first time, micro-arcsecond resolution of the nuclei of AG --- a factor of 30--100 greater resolution than is possible with ground-based millimeter VLBI.
\end{abstract}
\section{Introduction}
\begin{figure*}
\centering
\includegraphics[width=\linewidth]{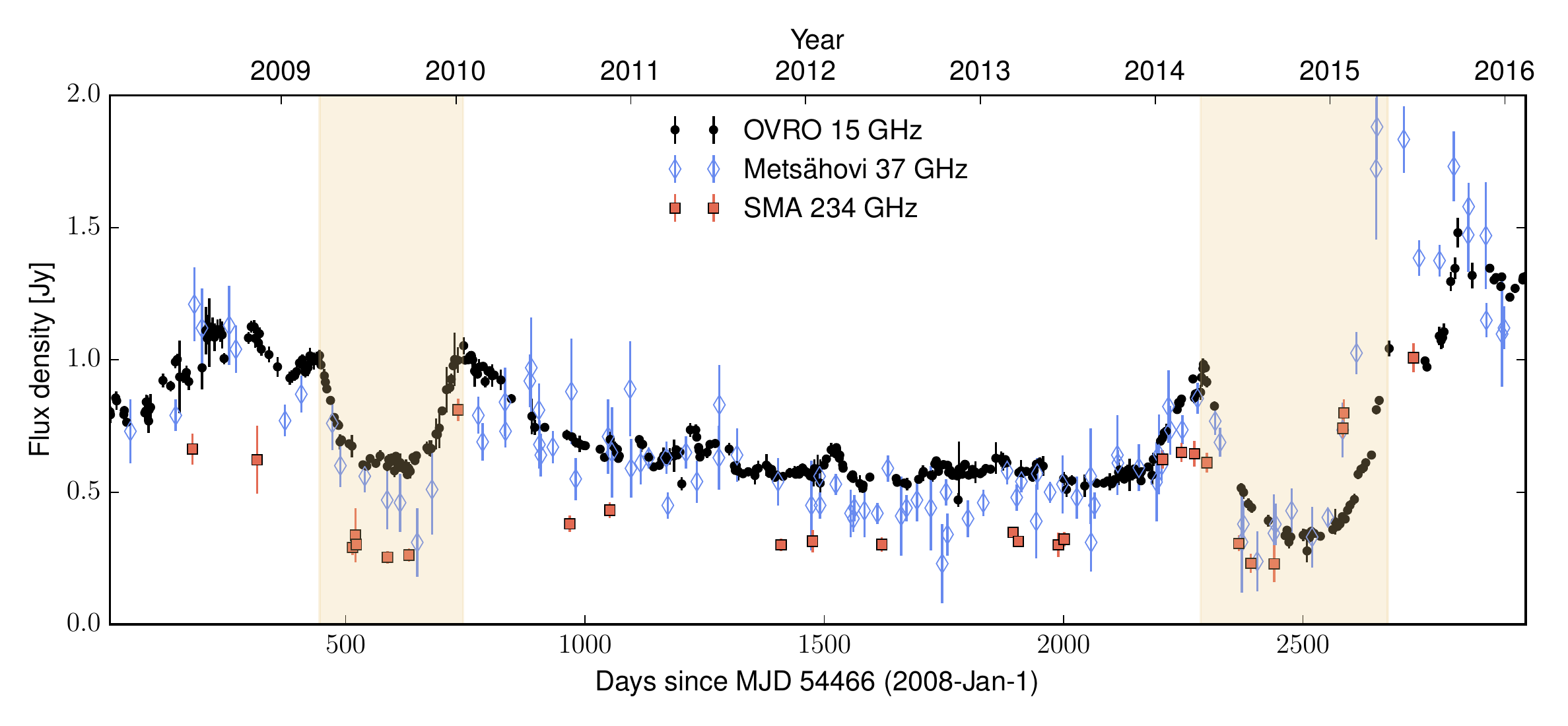}
\caption{Multi-frequency light curves of J1415+1320 showing symmetric achromatic variability (SAV). The roughly 1-year long U-shaped SAV events in 2009 and 2014 are highlighted.\label{fig:light_curve}}
\end{figure*}

We report a hitherto unrecognized form of radio variability in Active Galaxies (AG) --- Symmetric Achromatic Variability (SAV) --- which is time-symmetric and achromatic from 15\,GHz to 234\,GHz.\footnote{We use ``SAV'' to denote Symmetric Achromatic Variability, Symmetric Achromatic Variations, and Symmetric Achromatic Variable, relying on the context to make the usage clear.}
%and which may prove to be a powerful new probe of cosmological substructure.\footnote{We use ``SAV'' to denote Symmetric Achromatic Variability, Symmetric Achromatic Variations, and Symmetric Achromatic Variable, relying on the context to make the usage clear.}
SAV was first noticed in an unusual year-long symmetric U-shaped feature that appeared in the $15$\,GHz light-curve of  the BL Lac object J1415+1320 (PKS\,1413+135), in 2009--2010 (see Fig.~\ref{fig:light_curve}). The radio source lies either within or behind a spiral galaxy at redshift $z=0.247$ \citep{1981Natur.293..711B,1991MNRAS.249..742M,1992ApJ...400L..13C}. The U-shaped symmetry is similar to that observed in extreme scattering events (ESEs)  \citep{1987Natur.326..675F,2013A&A...555A..80P} caused by interstellar plasma structures, but the achromaticity, among other factors, can be shown to rule out ESEs as the origin of SAV \citep{j1415_apj_ese}. A pair of forward-reverse shocks within the source may create achromatic U-shaped events  \citep{2003ApJ...582L..75K}, but these are not expected to yield the high degree of symmetry observed in J1415+1320. %Furthermore, symmetry in nature is usually found to be rooted in deeper causes rather than random coincidences, as would be required were intrinsic fluctuations at the root of SAV. 

We propose that SAV is caused by the modulation of the flux density of a luminal or superluminal  \citep{1984RvMP...56..255B} compact emission region as it traverses the magnification pattern cast by an intervening gravitational lens. 
%Gravitational lensing can generate multiple magnified images of a background source and can provide a powerful  probe of the distribution of matter over cosmological distances.
The proposed lensing mechanism is akin to microlensing of stars by stellar-mass lenses \citep{1993Natur.365..621A}, but the lenses would need to be in the milli-lensing mass range $\sim
10^3$--$10^6\,{\rm M}_{\odot}$, with projected surface mass density of $\gtrsim 10^4\,$M$_\odot$\,pc$^{-2}$. Potential lens candidates with these properties include the dense cores of globular clusters and molecular clouds, and massive black holes. The lenses likely consist of multiple components that are projected close to the line of sight to the source, but are not necessarily gravitationally bound to each other. 

The intermediate mass range  has not been accessible by traditional centimeter-wavelength very-long-baseline interferometry (VLBI) searches for multiple images \citep{2001PhRvL..86..584W} or time delays \citep{1999ApJ...527..498F}, and although the upper half of this mass range is now being probed by millimeter wavelength VLBI \citep{2010arXiv1005.5551M, 2016MNRAS.462.1382B}, the lower half is still barely within reach. SAV may thus provide a powerful new method of exploiting gravitational lensing in addition to the two traditional methods employing multiple images and time delays. We note that our gravitational milli-lensing hypothesis leads to a number of testable predictions and is correspondingly fragile. We regard this as a strength of the hypothesis --- it should be easy to disprove.

The paper is organized as follows. \S 2 describes the Owens Valley Radio Observatory (OVRO) blazar monitoring survey in which SAV was first identified, and presents observational details of multi-frequency radio light curves of J1415+1320. In \S 3 we establish the statistical significance of the SAV features observed in J1415+1320, by quantifying the incidence of U-shaped features in the complete OVRO survey sample. \S 4 summarizes the known radio and optical properties of J1415+1320. In \S 5 we present simple lens models that can account for the observed SAV features in J1415+1320 as an ``existence proof.'' Though the light curves cannot be inverted to obtain a unique lens model, we exploit some generic properties of gravitational lenses to place constraints on the lens mass and projected density. Finally we discuss both the general and J1415+1320-specific implications of the lensing hypothesis in \S 6.

%We think that SAV is due to gravitational lenses in intervening galaxies along the line of sight that create complex caustic networks. A modest projected density of $\gtrsim 10^4 M_{\odot} {\rm pc}^{-2}$ is all that is required. Potential candidates include globular clusters, star clusters, molecular clouds, etc. Exotic lenses such as a super-massive black holes are not required. The lensing masses, need not be gravitationally bound --- all that is required is that this projected mass per unit area along the line of sight is contained within the intervening galaxy. 

%

%
\section{The Observations}
We report here on four series of radio flux density-monitoring observations, carried out at $15$\,GHz on the 40-m telescope at OVRO, at $22$\,GHz and $37$\,GHz on the 13.7-m telescope of the Mets\"ahovi Radio Observatory (MRO), at $100$\,GHz on the $6 \times 10.4$-m OVRO millimeter array, and at $234$\,GHz on the $8\times 6$-m Sub-Millimeter Array (SMA) of the Smithsonian Astrophysical Observatory (SAO).

\subsection{OVRO 15 GHz}
\subsubsection{The Sample}
Since 2008 the OVRO 40-m telescope blazar monitoring survey has accumulated $\sim 12,000$ object-years of observations and is by far the largest and most sensitive radio monitoring survey of blazars undertaken thus far.  The survey is being carried out in support of the \textit{Fermi} Gamma-Ray Space Telescope and the sample comprises mostly candidate gamma-ray blazars.  This program started as a survey of $15$\,GHz radio
variability of a complete sample of 1158 active galaxies 
\citep{2011ApJS..194...29R}, which has been augmented by the addition of several hundred \textit{Fermi}-detected blazars not in the original sample, samples of galactic micro-quasars, NLS1 galaxies, and the sample of sources detected at TeV energies,\footnote{\url{http://tevcat.uchicago.edu/}} so that  our whole sample now contains $\sim 1830$ objects observed at high-cadence --- twice-weekly, hardware problems and weather permitting. 

The complete sample comprising $1158$ objects was selected from a uniform all-sky survey of bright quasars in the Candidate Gamma-Ray Blazar Survey (CGRaBS) \citep{healey2008}. The CGRaBS sample was selected from objects in the Cosmic Lens All Sky Survey \citep{myers2003}, together with observations at both $4.8$\,GHz and $8.4$\,GHz. Over $11,000$ objects brighter than $65$\,mJy at $4.8$\,GHz, and with spectral index $\alpha > -0.5$ ($S \propto \nu^\alpha$) were selected initially. Since the 40-m program is in support of Fermi-GST, we restricted the sample to EGRET-like blazars \citep{hartman1999,mattox2001}, and so applied a figure of merit (FoM) based on 3EG blazar objects in CLASS, which yielded 1625 objects of which $1158$ have declination $>-20^\circ$  and can  be monitored by the 40-m telescope. We concentrate here on this statistically complete sample, in which we find that 981 objects have a high enough signal-to-noise ratio to identify symmetrical U-shaped features in their light curves.

\subsubsection{The Observations}
The 15\,GHz observations were made with the 40-m telescope of the Owens Valley Radio Observatory (OVRO) in California ($37.2314^\circ$\,N, 118.2827$^\circ$\,W). The 40-m telescope uses off-axis dual-beam optics and a cryogenic receiver with a 15\,GHz center frequency and 3\,GHz bandwidth. The two sky beams are Dicke-switched using the off-source beam as a reference. The source is alternated between the two beams in an ON-ON fashion to remove atmospheric and ground contamination. The on-source observing time is $32$\,s and the $3\sigma$ detection limit is $12$\,mJy in good weather.

In May 2014, a new pseudo-correlation receiver was installed on the 40-m telescope, and the gain variations have since been corrected using a 180-degree phase switch instead of a Dicke switch. The performance of the new receiver is very similar to the old one and no discontinuity is seen in the light curves. Calibration is achieved using a temperature-stable diode noise source to remove receiver gain drifts, and the flux density scale is derived from observations of 3C\,286 assuming a value of 3.44\,Jy at 15\,GHz \citep{baars1977}. The systematic uncertainty of about 5\% in the flux density scale is not included in the error bars. Further details of the reduction and calibration procedures may be found in \cite{2011ApJS..194...29R}.

\subsection{Mets\"{a}hovi 22 and 37\,GHz}
The $22$ and $37$\,GHz observations were made with the 13.7-m diameter Aalto University Mets\"ahovi radio telescope, which is a radome-enclosed Cassegrain antenna in Finland ($60.218^\circ$\,N, $24.393^\circ$\,E). The measurements were made with $1$\,GHz-band dual-beam receivers centered at $22.2$ and $36.8$\,GHz respectively. The HEMT (high electron mobility transistor) front end operates at room temperature. The observations are Dicke-switched ON--ON observations, alternating the source and an adjacent patch of sky in each feed horn. The typical integration time for one flux-density measurement is $1200$--$1600$\,s. The detection limit is about $0.2$\,Jy under optimal conditions. Data points with a signal-to-noise ratio $< 4$ are considered non-detections. The flux-density scale is set by observations of the H{\sc ii} region DR\,21. Sources NGC\,7027, 3C\,274 and 3C\,84 are used as secondary calibrators. A detailed description of the data reduction and analysis is given in \cite{metsahovi}. The error estimate in the flux density includes the contribution from the measurement rms and the uncertainty of the absolute calibration. 

\subsection{OVRO 100 GHz}

Flux densities in the 3\,mm band (100\,GHz) were obtained in the period 1992--2003 from the flux density history catalog of the Owens Valley Radio Observatory Millimeter Array (MMA), a six-element millimeter wave radio telescope array located at the main OVRO site until 2005. The observations  were primarily made during dedicated calibration sessions to track the flux density of millimeter-loud blazars and other point sources. The flux-density scale was  calibrated using contemporaneous measurements of Uranus and/or Neptune.

\subsection{SMA 234 GHz}
Flux densities  were measured at the Submillimeter Array (SMA), an 8-element interferometer located near the summit of Mauna Kea (Hawaii). 
The measurements were made at frequencies ranging from 221 to 297 GHz; we use the mean frequency of 234 GHz in our analysis. J1415+1320 is included in an ongoing monitoring program at the SMA to determine the flux densities of compact extragalactic radio sources that can be used as calibrators at millimeter and submillimeter wavelengths \citep{2007ASPC..375..234G}.  Potential calibrators are from time to time observed for 3 to 5 minutes, and calibrated against known standards, typically solar system objects (Titan, Uranus, Neptune, or Callisto).  Data from this program are updated regularly and are available at the SMA website.\footnote{\url{http://sma1.sma.hawaii.edu/callist/callist.html}}

\section{Symmetric Achromatic Variability (SAV)}
\begin{figure*}
\centering
\includegraphics[width=\linewidth]{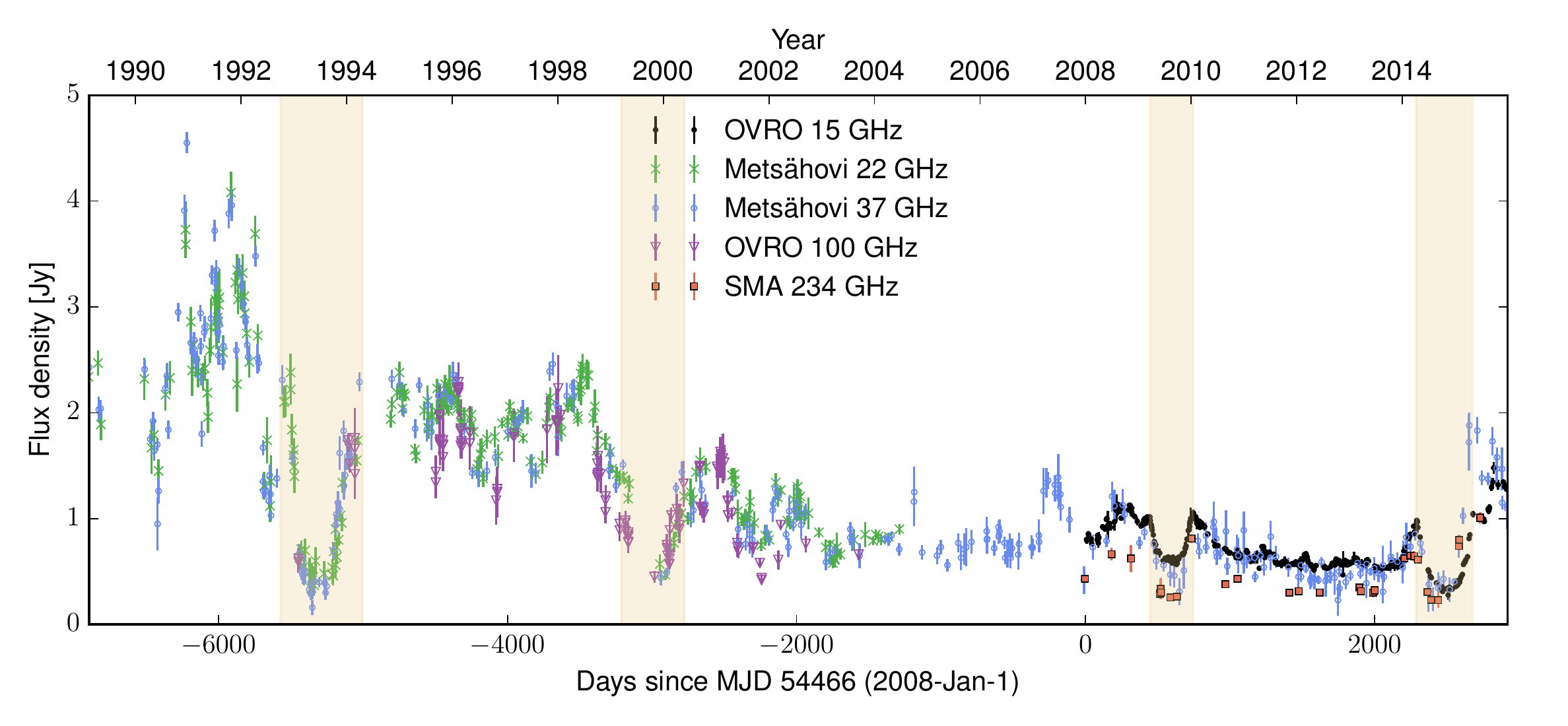}
\caption{Same as Fig.~\ref{fig:light_curve} but with the entire Mets\"{a}hovi monitoring data at 22 and 37\,GHz since 1989, and the OVRO MMA data at 100\,GHz. Two more achromatic U-shaped events are seen in 1993 and 2000.\label{fig:light_curve_full}}
\end{figure*}

In 2009, we observed a roughly 1-year-long highly symmetric U-shaped feature in the OVRO light curve of J1415+1320 and a similar feature was seen in 2014. Fig.~\ref{fig:light_curve} shows the 15\,GHz (OVRO), 37\,GHz (Mets\"ahovi), and 234\,GHz (SAO) light curves. The U-shaped features are highly achromatic from 15\,GHz to 234\,GHz, although there are
small deviations from perfect achromaticism. We therefore refer to this type of variability as ``Symmetric Achromatic Variability,'' or SAV. %Since gravitational lensing is always achromatic and for simple lenses it is also generally symmetric this seemed a likely explanation for SAV.
The object has also been monitored at 22\,GHz and 37\,GHz at  Mets\"ahovi since 1989, and two more U-shaped features were observed in 1993 and 2000, which were also observed on the OVRO MMA at 100\,GHz (Fig.~\ref{fig:light_curve_full}). Thus these two additional U-shaped features are also achromatic over the whole observed range --- here from 22\,GHz to 100\,GHz.

The details of the profile of the U-shaped features are seen most clearly in the high-cadence, high-sensitivity OVRO observations from 2008--2016 (Fig.~\ref{fig:light_curve}). These show a distinctive mirror symmetry composed of a pair of spikes with time-reversed profiles.  As described in \S 5 in the context of gravitational lensing, we can distinguish two types of U-shaped events: a ``volcano'' type in which the first spike displays a slow rise and a fast decline and the second spike shows a fast rise and slow decline (SRFD--FRSD); and a ``crater'' type in which a fast rise, slow decline is followed by a slow rise, fast decline (FRSD--SRFD). All four events seen in J1415+1320 are of the volcano type. We see from Figs.~\ref{fig:light_curve} and \ref{fig:light_curve_full} that an eight-year period of unusually low activity has enabled us to detect SAV clearly in J1415+1320 between 2008 and 2016.

\subsection{Incidence of U-shaped events}
\begin{figure*}
\centering
\includegraphics[width=0.48\linewidth]{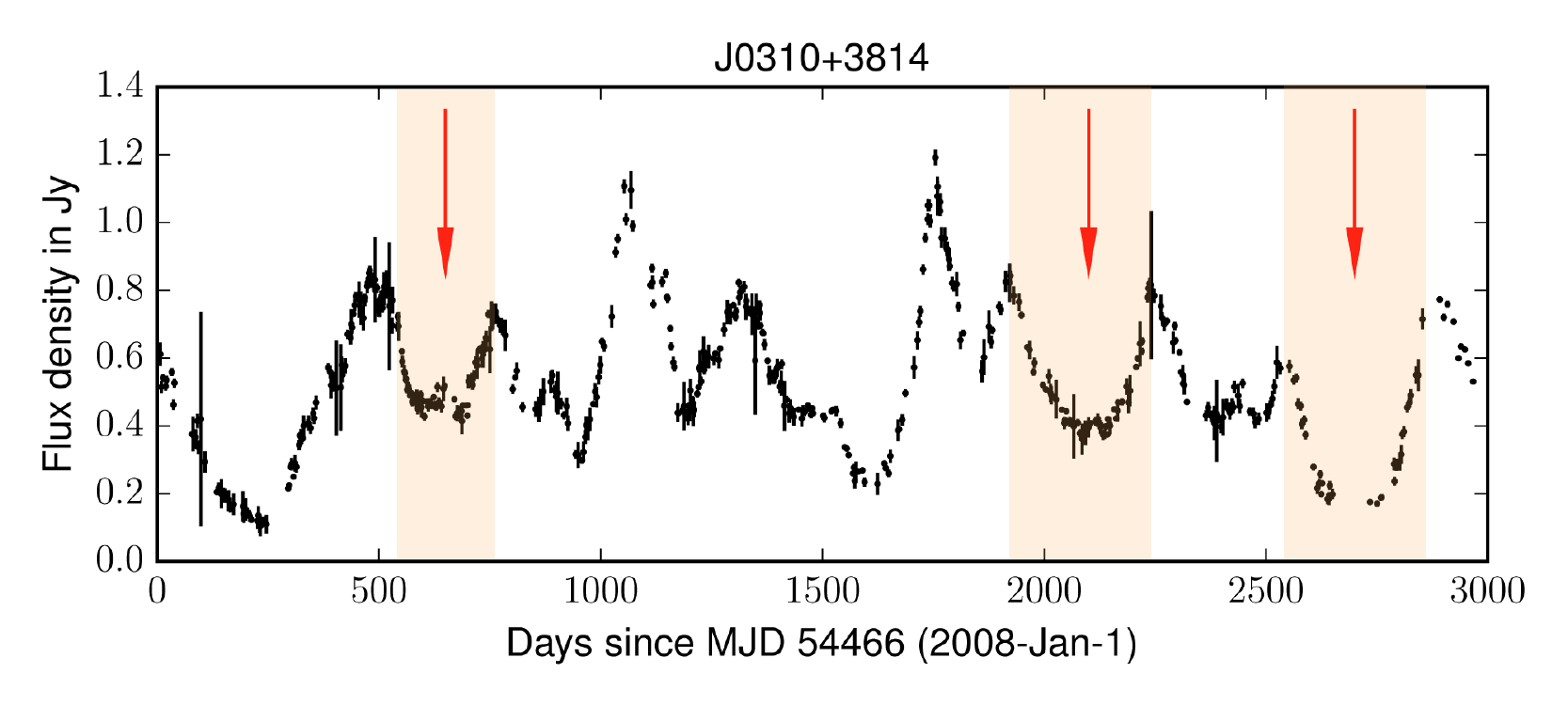}
\includegraphics[width=0.48\linewidth]{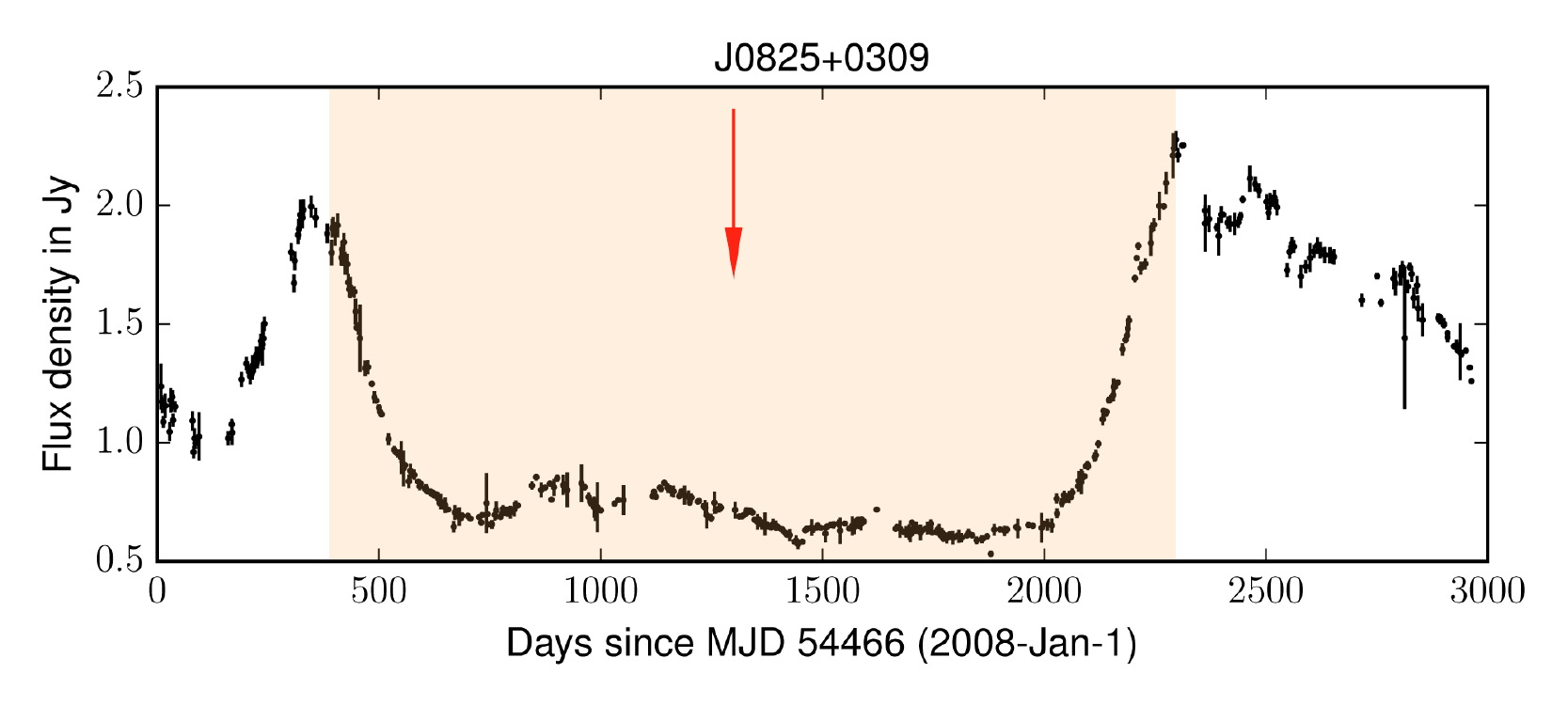}\\
\includegraphics[width=0.48\linewidth]{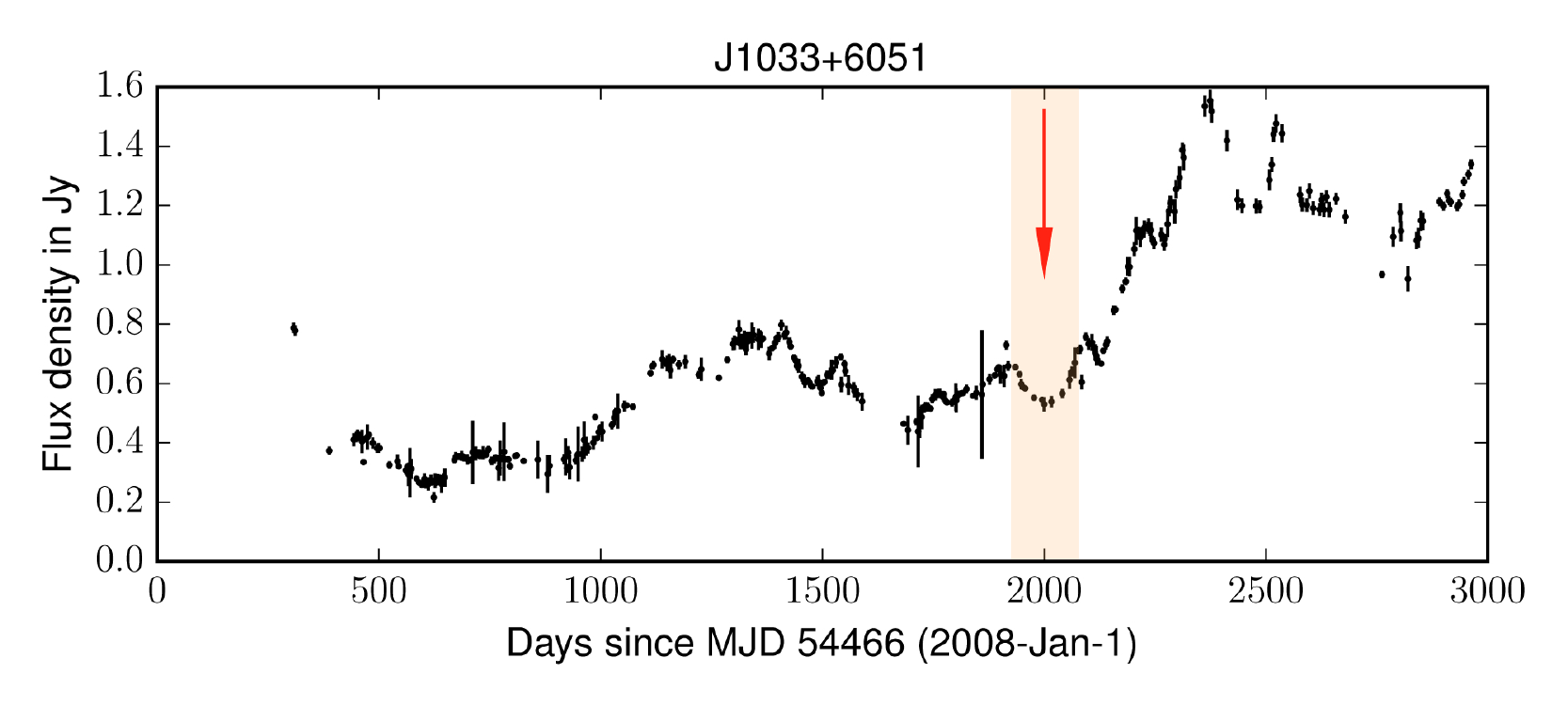}
\includegraphics[width=0.48\linewidth]{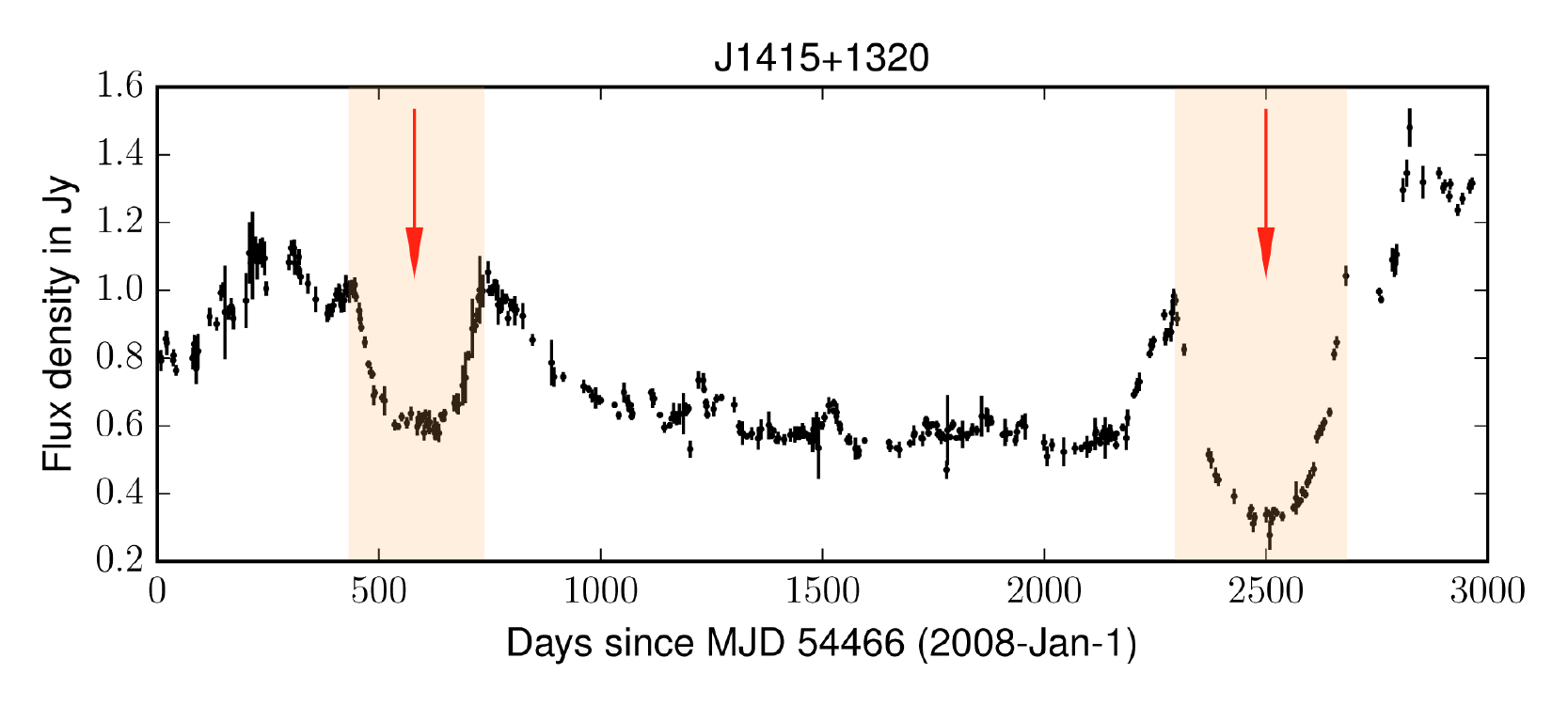}\\
\includegraphics[width=0.48\linewidth]{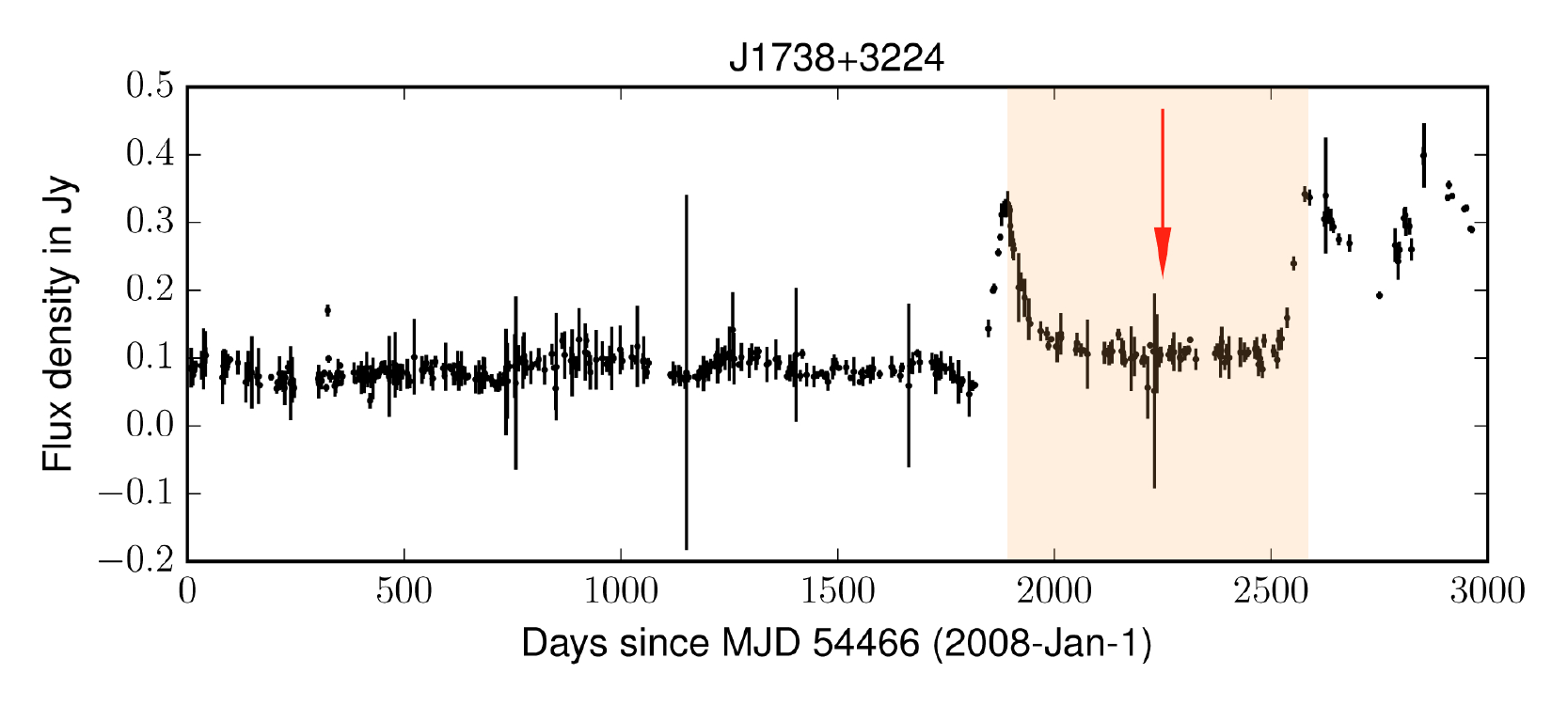}
\includegraphics[width=0.48\linewidth]{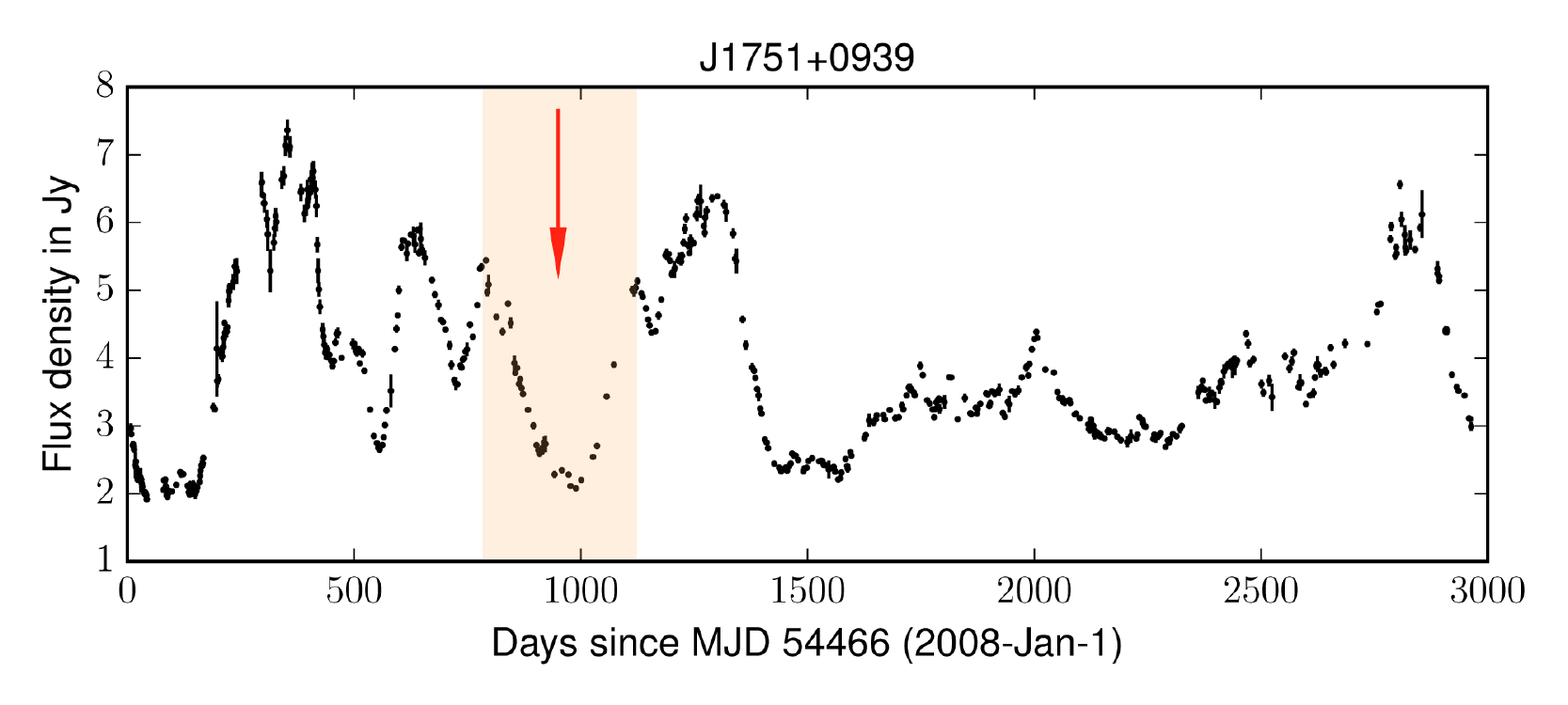}\\
\includegraphics[width=0.48\linewidth]{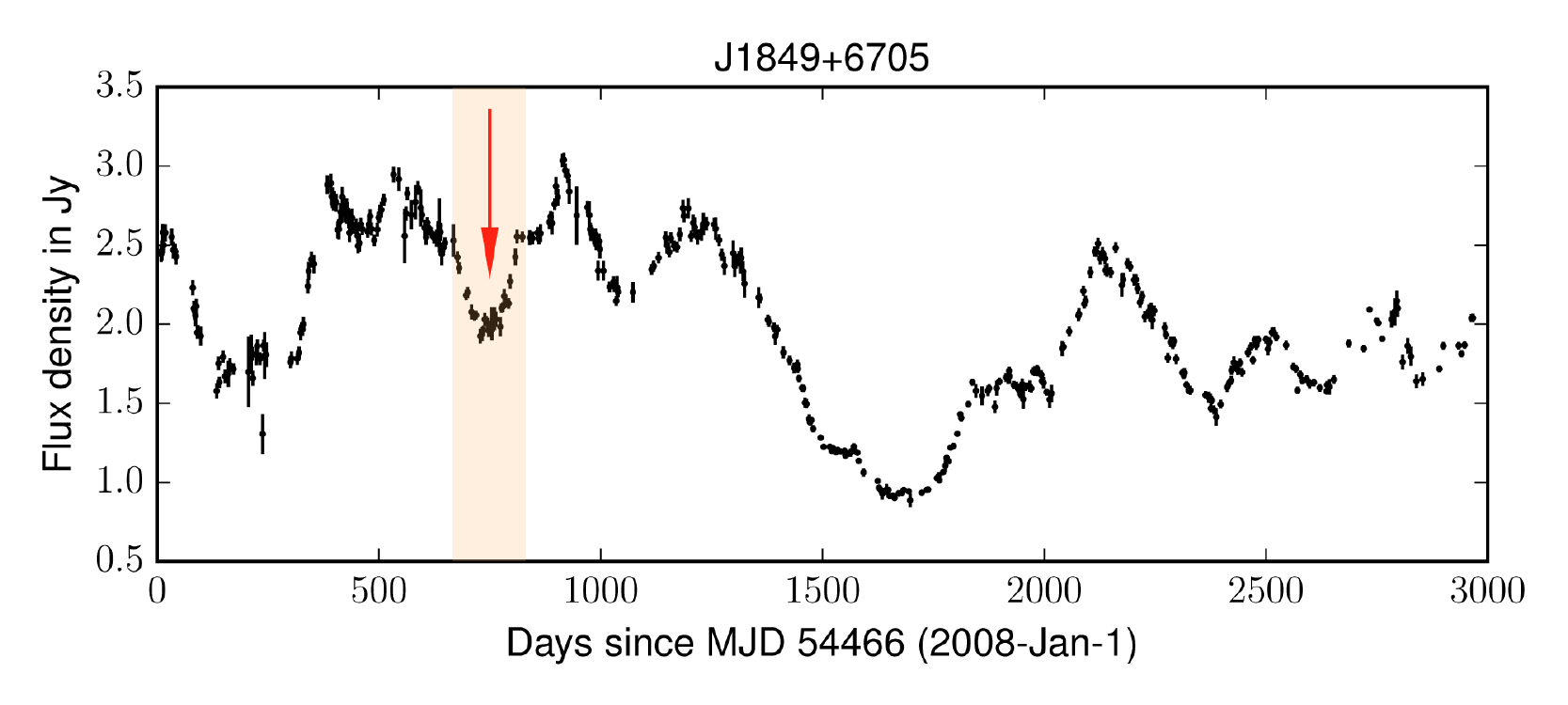}
\caption{OVRO 15\,GHz light curves of seven objects in which U-shaped features were detected. The ten features are indicated with arrows.\label{fig:sav_candidates}}
\end{figure*}
Many sources in the complete sample often show ``flares'' that can have either 
SRFD or FRSD shape \citep{2008A&A...485...51H}. To test if the observed U-shaped events are due to a chance juxtaposition of two flares, we examined the light-curves of
the complete sample from 2008--2016 for both volcano and crater type U-shaped events.
%To study the statistical significance of the 2009 U-shaped feature in J1415+1320, we looked for more examples in the core sample. 
%Ideally one would like to employ digital fitting and machine learning, and we have embarked on such a program. In the meantime we have adopted a pragmatic approach. 
Two of the co-authors examined the $981$ high-quality light curves independently and selected all those U-shaped features that they deemed to be of sufficient symmetry to be of interest. 
%We discarded $177$ of the light-curves that are not of high enough quality to pick out U-shaped features, reducing the core sample to $981$ objects. 
One co-author picked out $23$ U-shaped features, and the other $25$ U-shaped features.  Only the ten features selected by both authors were accepted as candidates. These features were seen in seven objects, of which one had two features and another had three (Fig.~\ref{fig:sav_candidates}). Thus U-shaped features in the OVRO light-curves are rare, and the random probability of a U-shaped feature in one of our $981$ 8-year light-curves is 1.02\%. Among these seven objects, J1415+1320 stands out in view of the clarity and isolation of the two U-shaped features relative to the rest of the light-curves. 
 If such U-shaped features are distributed randomly among all the objects, the probability of three or more occurrences in one object in 8 years (J0310+3814) is $10^{-6}$, while the probability of three or more occurrences in one object in 27 years (J1415+1320) is $\sim 4 \times 10^{-6}$. It is thus highly likely that the U-shaped features we have observed in J0310+3814 and J1415+1320 are not random events but are an unusual intrinsic feature of the objects themselves, are due to  propagation effects along the lines of sight, or are a combination of both. At this stage SAV has not been demonstrated in J0310+3814, since we have no proof that the symmetric variations are achromatic.

\begin{figure*}
\centering
\includegraphics[width=\linewidth]{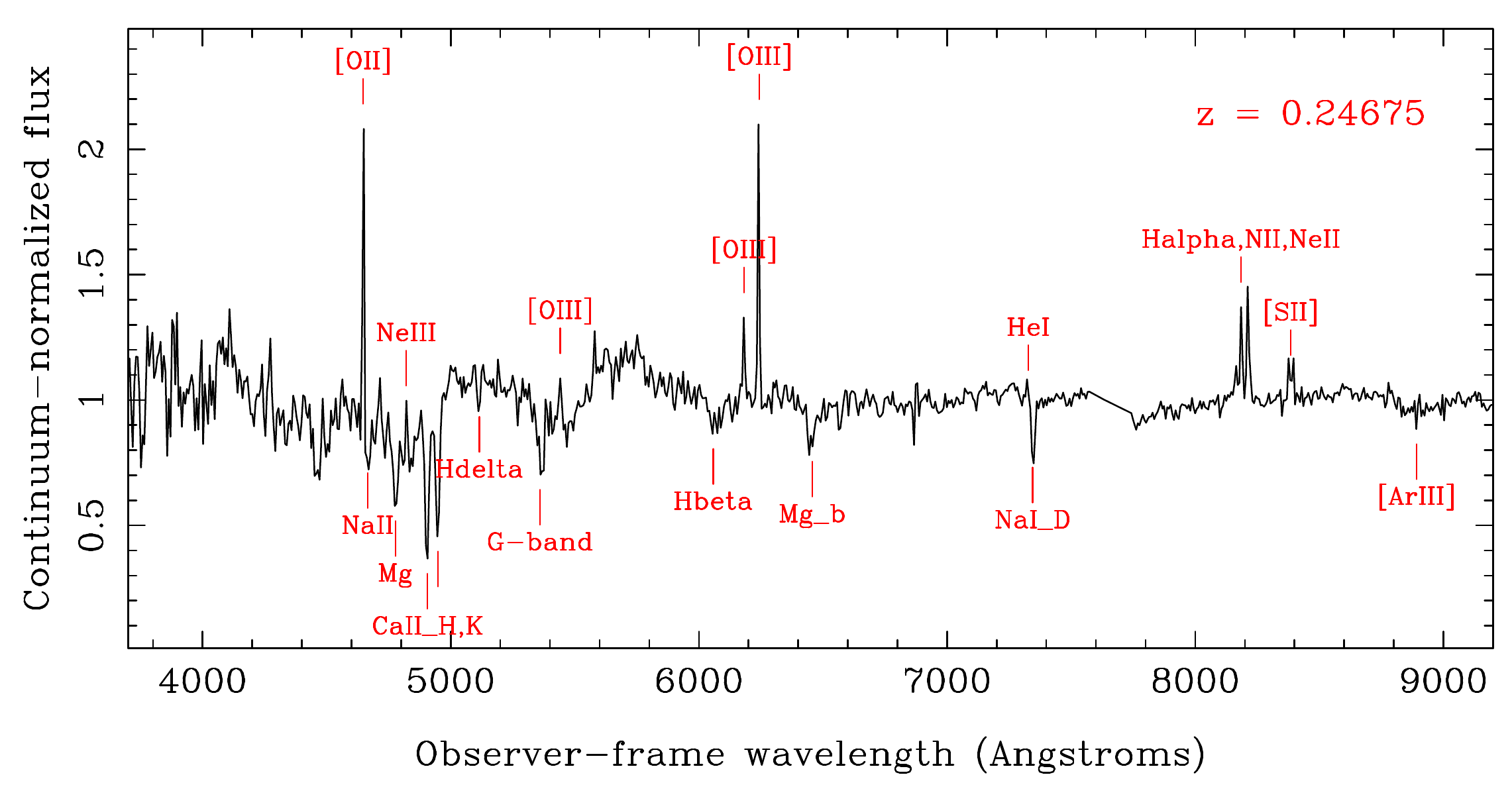}
\caption{Continuum normalized LRIS spectrum of the putative host of J1415+1320. Prominent lines are marked for a redshift of $z=0.24675$, which agrees with that determined from \ion{H}{1} 21-cm absorption. There is no evidence for emission from an intervening redshift. The artifacts near 6877\,\AA\  and 7613\,\AA\ are due to imperfect subtraction of sky-lines.\label{fig:lris}}
\end{figure*}

\section{Radio and optical properties of J1415+1320}
J1415+1320 has a two-sided jet and exhibits up to eight mas-scale components extending over $\sim60$\,mas, some of which move outward with speeds ranging from $0.06$--$0.3$\,mas\,yr$^{-1}$ and apparent
transverse speeds up to $1.5c$  \citep{1998AJ....115.1295K,2002AJ....124.2401P,2005ApJ...622..136G,2009AJ....138.1874L,2016AJ....152...12L}. Based on its radio morphology, J1415+1320 is classified as a Compact Symmetric Object (CSO), with radio lobes straddling a bright core \citep{1992ApJ...396...62C,1994ApJ...432L..87W,1996ApJ...460..612R}. Although the cores in CSOs are bright \citep{1992ApJ...396...62C} and variable, the core of J1415+1320 is
exceptional -- with a core fraction $>0.5$ \citep{2009AJ....138.1874L} it is both far brighter and far more variable than most CSO cores in radio galaxies, which  have core fractions in the
range $\sim 0.004$--$0.06$ \citep{1996ApJ...463...95T}. In addition, J1415+1320 is identified with a spiral galaxy at $z=0.247$ viewed edge-on \citep{1981Natur.293..711B,1991MNRAS.249..742M,1992ApJ...400L..13C}, whereas BL-Lac objects are rarely, if ever, hosted by spiral galaxies. The radio core is offset from the optical center of the spiral galaxy by  0.013\,arcsec (50\,pc) \citep{2002AJ....124.2401P}. These unusual properties, amongst others,  have led to suggestions that the radio source is an unrelated background object that is being lensed by matter in the spiral galaxy \citep{1992ApJ...400L..13C, 1992ApJ...400L..17S}. We note that based on the number densities of radio sources and intervening galaxies, the probability of such a chance alignment is small ($\approx 0.006$; details of calculation in Appendix~\ref{sec:chance_alignment}). However, this probability is based on {\it a posteriori} statistics and as such the background-source hypothesis cannot be ruled out. If the radio source is a background object, then due to the lack of multiple images on arcsecond scales (due to lensing by the bulge of the spiral), the redshift of the radio source is bounded by $z\lesssim 0.5$ \citep{1999MNRAS.309.1085L}.

\subsection{Optical spectrum of J1415+1320}
We obtained an optical spectrum of the spiral galaxy with the low resolution imaging spectrometer (LRIS) \citep{lris} on the Keck telescope. The data were obtained on MJD\,57488.43 with a total exposure time of about 1\,hour. The \texttt{lpipe}\footnote{\url{http://www.astro.caltech.edu/~dperley/programs/lpipe.html}} software was used for bias subtraction, flat fielding, cosmic-ray rejection, and sky-line subtraction. We then obtained a wavelength solution using arc-lamps and sky lines, optimally extracted the spectrum, and calibrated the flux-density and atmospheric opacity. The resulting spectrum is shown in  Fig.~\ref{fig:lris}. We detected several emission and absorption lines with high significance, all of which can be attributed to the spiral galaxy at a redshift consistent with that obtained from 21-cm absorption \citep{1992ApJ...400L..13C}. The spectrum also contains some lines (not annotated in Fig.~\ref{fig:lris}) detected at low significance which are consistent with $z=0$ and likely caused by Galactic absorption. We find no clear evidence for an intervening galaxy. We also did not detect broad lines from the AGN, owing to severe extinction by foreground dust-lanes in the spiral galaxy. As such, the redshift of the AGN could not be independently determined.

\section{SAVs from Gravitational Lensing}
As shown in \S 3, the statistics of occurrence of symmetric U-shaped events makes them highly unlikely to be caused by random intrinsic fluctuations.  In a companion paper \citep{j1415_apj_ese}, we have studied in detail the possibility that SAV could be caused by ESEs and shown that this explanation for the SAV phenomenon can be ruled out. In this section, we show that the SAVs we observe in J1415+1320 can be explained naturally and economically by gravitational lensing.

Gravitational lensing is frequency-independent and accounts for the achromatic nature of SAVs. Relatively simple lens configurations yield symmetric magnification patterns. Time-dependent magnification occurs as a source moves over a magnification pattern cast by a foreground lens mass. The most prominent magnification features expected from gravitational 
lensing are fold caustics \citep{1986ApJ...310..568B,1988PhRvA..38.4028B,1992ARA&A..30..311B}. Fold caustics are curves in the source plane where the point source magnification diverges.\footnote{If we allow the distance of the source to vary, they should be seen as surfaces.} Rapid brightness changes ensue as a source crosses a fold caustic, and U-shaped features are a result of a closely spaced pair of fold crossings. Such U-shaped events, due to their symmetry, are a tell-tale sign of lensing, and a powerful discriminant between intrinsic and lensing-induced variability \citep{glbook}.

The generic magnification pattern in gravitational lensing comprises fold curves meeting at cusp points. When a source crosses into a closed fold contour at time $t_c$, the flux variation shows a rapid rise with a relatively slow decay that follows a $|t-t_c|^{-1/2}$ law \citep{glbook}. This yields an FRSD feature. Source motion in the opposite direction naturally yields an SRFD curve. Crater symmetries (FRSD--SRFD) occur when a source crosses two folds that meet at a nearby cusp. Volcano symmetries (SRFD--FRSD) occur when a source crosses out of a closed fold contour and subsequently crosses back into it. Successive U-shaped SAV features may or may not come from the same source component; if they come from successive components following a similar track, then we expect the time series to be similar, although the transverse speed may be different leading to a temporal scaling of the flux density variations.

\subsection{An existence proof}
\begin{figure*}
\centering
\includegraphics[width=\linewidth]{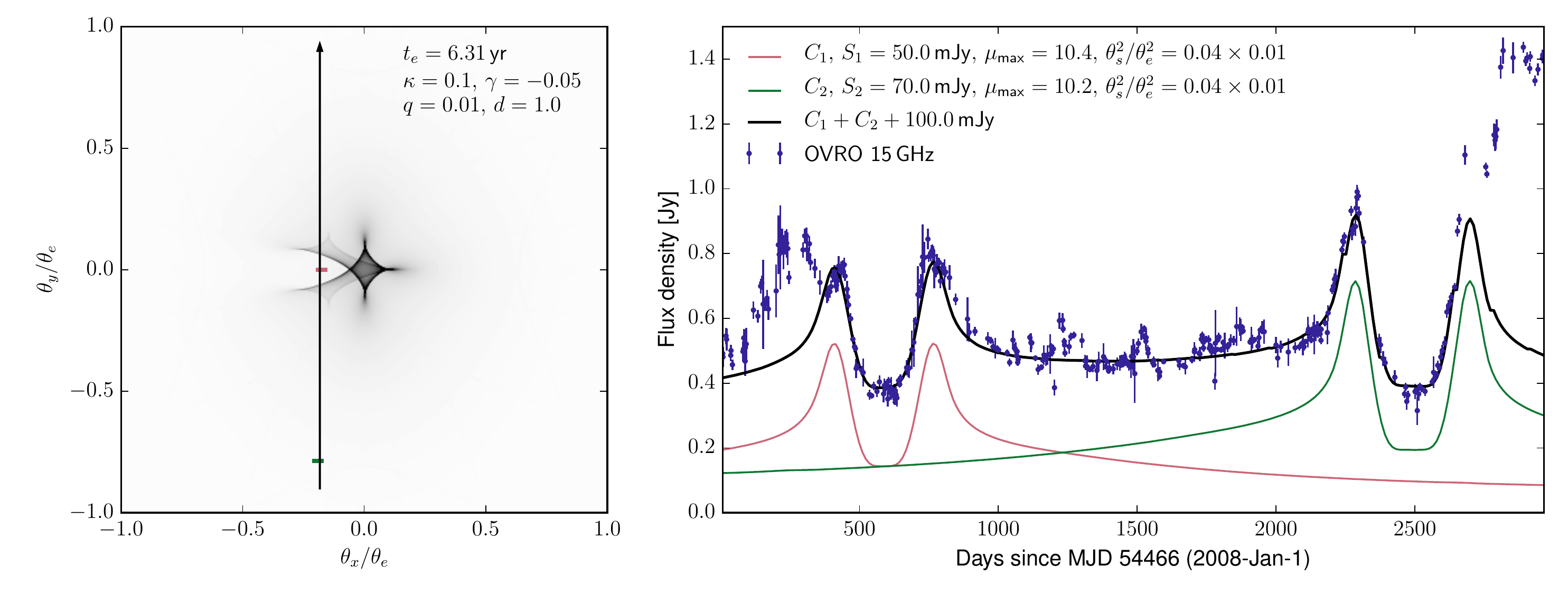}
\includegraphics[width=\linewidth]{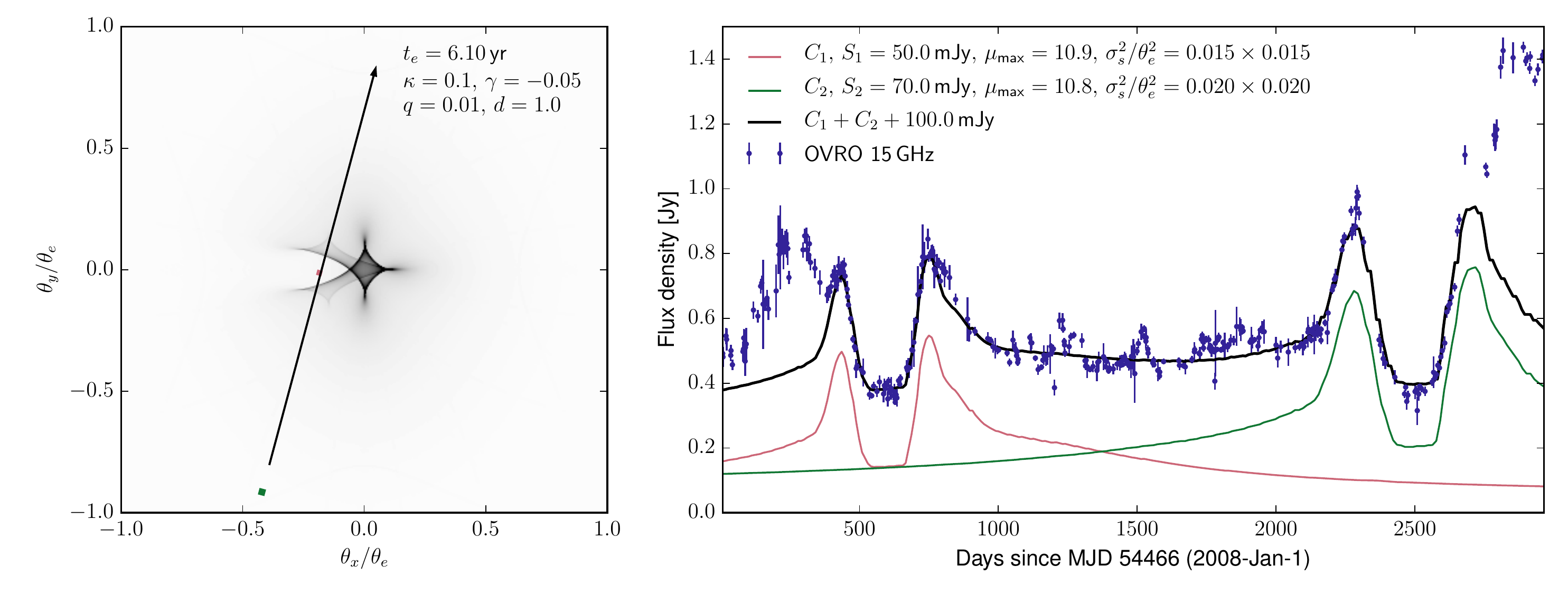}
\includegraphics[width=\linewidth]{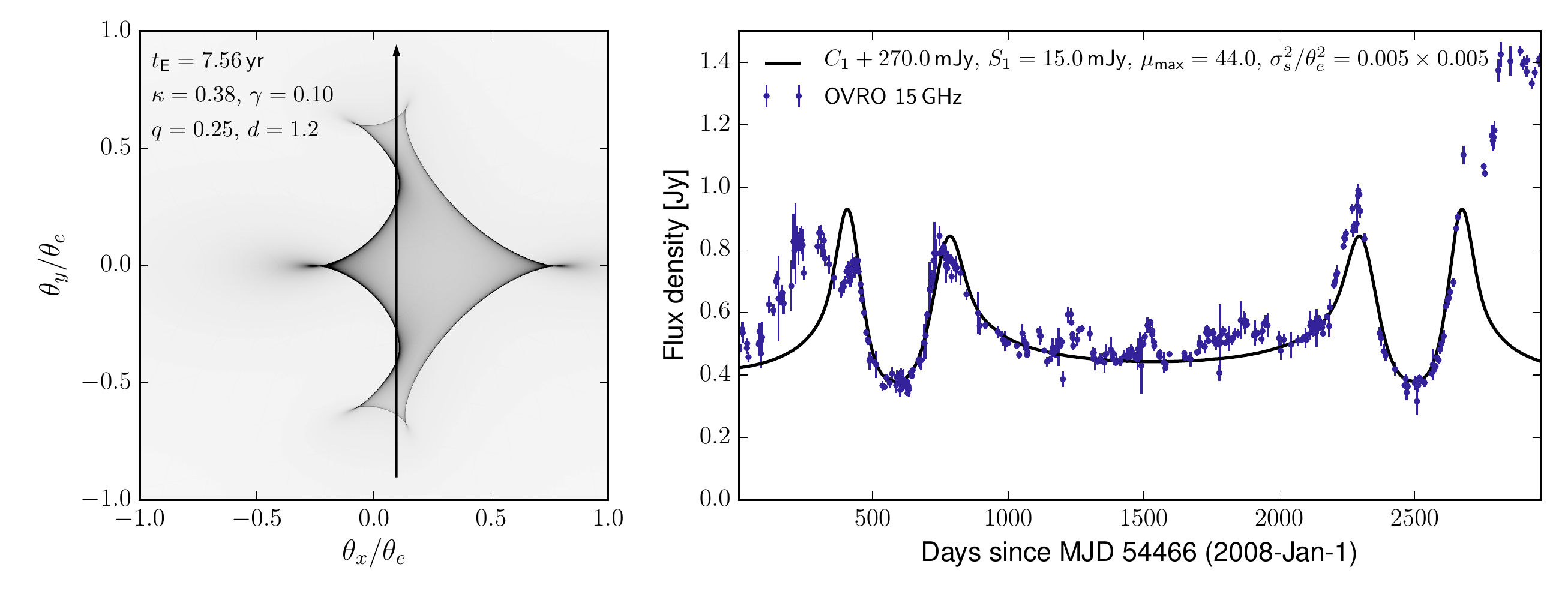}
\caption{Fits to the OVRO 15\,GHz data for two lens models. A linear function of time has been subtracted from the data.
Left panels show the magnification pattern, and right panels show the model light curves. Top two rows have the same lens model but different trajectories for the two source components (red and green). Right panels show the light curve for the individual components (green and red) and their sum (black). The lens model in the bottom-most row requires only one source components but the range of plausible trajectories is limited. Lens and source parameter values are shown in the embedded text.\label{fig:model_fits}}
\end{figure*}

To establish a ``proof of existence'' for the gravitational lensing hypothesis, we explored various commonly encountered lens configurations that can yield the basic unit of SAV -- a U-shaped event with volcano (SRFD--FRSD) symmetry -- as a result of rectilinear motion of the source. Lenses with a high degree of circular symmetry typically yield a four-cusped diamond-shaped caustic configuration (called an astroid). Barring contrived source trajectories, such a lens yields crater-type light curves. The simplest asymmetric lens configuration is a binary lens (the binary need not be gravitationally bound). Binary lenses typically yield a six-cusped caustic pattern which has sufficient complexity to generate a large range of light curve features including both crater and volcano type SAVs.

We simulated several binary lenses by varying the binary separation $d$ and mass-ratio $q<1$ and found that a range of binary lens parameters can yield the observed U-shaped features for a variety of source trajectories (details of the simulations are presented in Appendix~\ref{sec:model_details}). We show two examples in Fig.~\ref{fig:model_fits}.  The upper two panels (left) show the magnification pattern of a binary lens with mass ratio of $q=0.01$ and separation of $d=\theta_{\rm E}$ ($\theta_{\rm E}$ is the Einstein radius of the more massive object). Since such binary condensates typically exist in the presence of a smoothly varying background mass distribution (such as in a galaxy), we added a background convergence  $\kappa=0.1$ and shear  $\gamma=-0.05$. These values are not very constraining, but are chosen to be ``sub-critical,'' in that the external mass distribution cannot strongly lens the source independently. An approximately south--north source trajectory across the lens (black arrows) generates a single SRFD--FRSD U-shaped event. Hence two compact source components can generate the entire 8\,year light curve to good accuracy. Note that we have removed a linear trend from the 15\,GHz data points to account for any intrinsic variability of the lensed components or that of unlensed source components away from the caustics.

The bottom panel of Fig.~\ref{fig:model_fits} shows an alternative binary model that can reproduce both U-shaped features with only one moving radio component rather than two. This however requires the source to ``graze'' the caustic pattern making the set of plausible source trajectories somewhat restrictive.  

The above fits were obtained by manually varying the lens and source parameters. In particular, the crossing time-scale, and hence the source speed, is largely set by the relative timing of the four peaks that comprise the two U-shaped events in 2009 and 2014. The source components' sizes and flux densities are largely determined by the peak-magnification at the caustics. These two factors are somewhat degenerate, because the magnification near a fold caustic is expected to follow $\mu=\alpha p^{-1/2}$, where $p=\sigma_s/\theta_{\rm E}$ where the source is modeled as a Gaussian with standard deviation $\sigma_s$, and $\alpha\approx 1$  depends on the precise nature of the Fermat potential at the fold. Hence a weaker smaller source can generate a caustic peak with a similar flux density as a stronger larger source. However, by exploring a large parameter range, we found that the source components must be smaller than $\approx 0.03\,\theta_{\rm E}$ to account for the precise light curve {\em profile} around the caustic crossing. For the family of lens model considered here we found $2.5\lesssim \alpha\lesssim 4.25$. Table~1 summarizes the parameter values for two lens models presented in Fig.~\ref{fig:model_fits}.

\begin{figure*}
\centering
\includegraphics[width=0.8\linewidth]{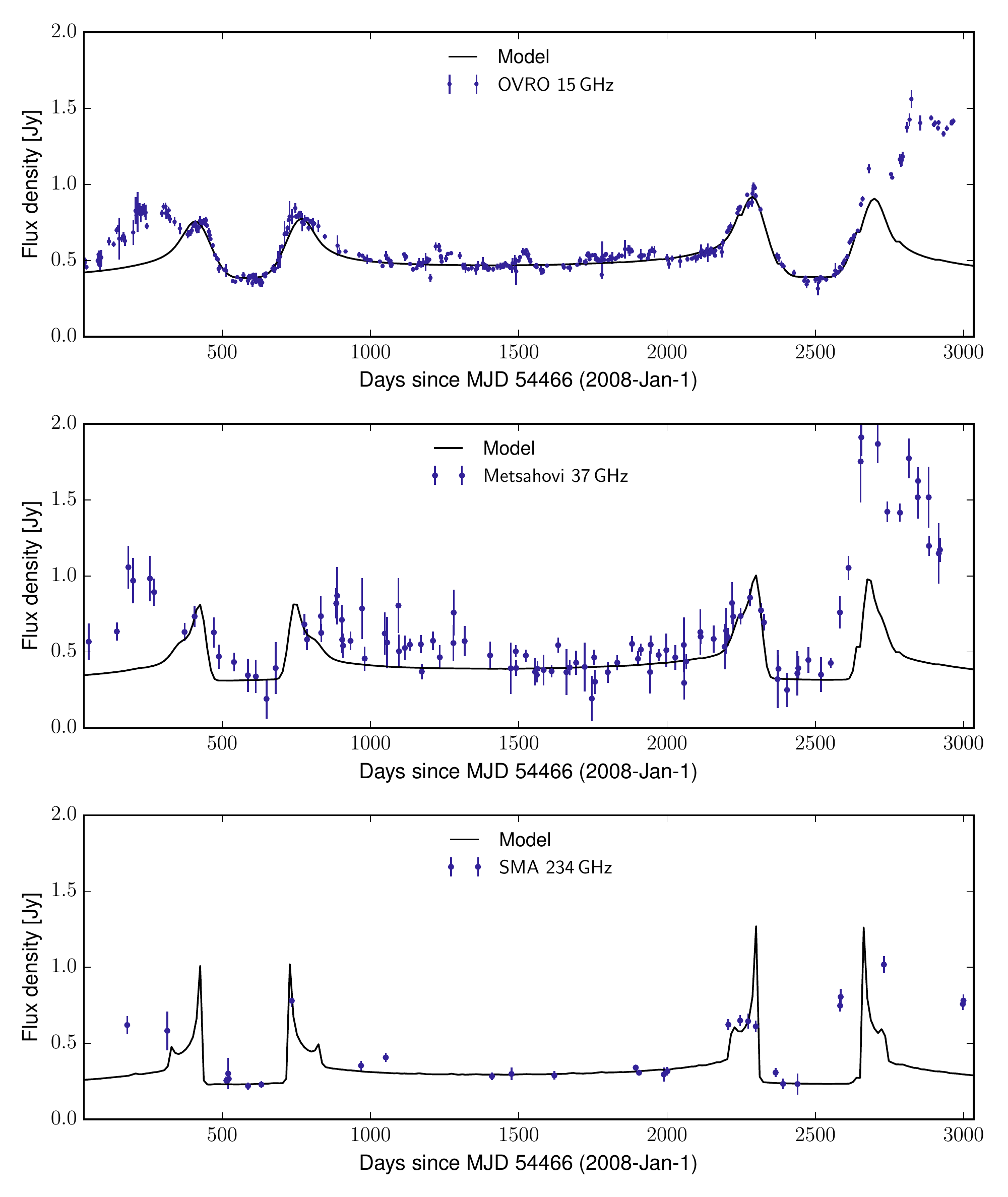}
\caption{Multifrequency model light curves over plotted on data. at 15, 37, and 234\,GHz. The lens model and source trajectories are the same as that in the top-most row of Fig. \ref{fig:model_fits}\label{fig:multifreq_fits}.}
\end{figure*}

A hallmark of gravitational lensing is its achromaticity, although we expect some chromaticity due to variation of source size and centroid position with frequency. Such variation is expected in the emission from relativistic jets due to a combination of the evolution of the emitting population of relativistic electrons with time and location in the jet, and synchrotron opacity effects. To test if the binary lens models reproduce the 37\,GHz and 234\,GHz observations, we assumed that (i) the source component size varies with frequency $\nu$ as $\nu^{-1}$, (ii) the flux densities of the unlensed source components have a spectral index of $\alpha=-0.7$, (iii) the lensed components have a flatter spectral index of $\alpha=-0.1$. With these assumptions, we computed the corresponding light curves at 37\,GHz and 234\,GHz where data are available. Fig.~\ref{fig:multifreq_fits} shows the multi-frequency data and the expected light curves for the lend model shown in the upper row of Fig.~\ref{fig:model_fits}. Despite simplistic assumptions on the source properties, the fits are quite good over a factor of $15$ in wavelength.

The data  show small but significant departures from both model light curves which may be due to intrinsic variability or due to our simplistic assumptions of both the source and lens. For instance, there is evidence for additional symmetries in the light curves (see Appendix~\ref{sec:extended_caustics}), which may be accounted for by lenses with a larger number of mass condensates that give rise to caustic networks with higher complexity \citep{1988PhRvA..38.4028B,1987A&A...171...49S}. Nevertheless, the models satisfactorily account for the symmetry and profile of the SAV features and serve as an ``existence proof'' for the gravitational lensing hypothesis.

\begin{table*}
\centering
\caption{Summary of lens and source parameters used in the model fits from Fig.~\ref{fig:model_fits}. The columns represent the binary mass ratio $q$, binary separation $d$, external convergence $\kappa$, external shear $\gamma$, Einstein crossing time $t_{\rm E}$, Einstein radius $\theta_{\rm E}$, transverse speed of the source $v_\perp$, and source size $p$. The Einstein angle assumes a source brightness temperature of $T_b=10^{13}$\,K, and the transverse speed values are based on a nominal source redshift of $z_s=0.5$ (cosmological time dilation included)}
\begin{tabular}{llllllll}
\hline\\
$q$     & $d/\theta_e$  & $\kappa$    & $\gamma$    & $t_{\rm E}$\,[yr]   & $\theta_{\rm E}\,[\mu$as]   & $v_\perp/c$   & $p=\sigma_s/\theta_{\rm E}$ \\\hline \\
0.25    & 1.2           & 0.38          & 0.1           & 7.54          & 81                    & 0.33          & $0.01\times 0.01$\\
%0.1     & 1.3           & 0.38          & -0.1          & 3.3           & 50.4                  & 0.47          & $0.03\times0.03$ \\
0.01    & 1.0           & 0.1           & -0.05         & 6.31          & 87.3                  & 0.42          & $0.04\times0.01$ \\ \hline
\end{tabular}
\end{table*}

\subsection{Constraints on lens properties}
Having seen that simple lens models can produce the observed SAV features in J1415+1320, we now place constraints on the physical properties of the putative lens. Although several lens configurations are viable, fold caustics have generic properties that enable us to constrain the lens parameters almost independently of the precise lens configuration.

We constrain the lens mass, $M$, by placing bounds on the Einstein radius, $\theta_{\rm E}$, that sets the natural length and angular scale for lensing \citep{1992ARA&A..30..311B}: 
\begin{eqnarray}
\theta_{\rm E} &=& (4GM/c^2D)^{1/2}\nonumber \\
&\approx& 90\,(M/10^3\,M_{\odot})^{1/2}(D/1\,{\rm Gpc})^{-1/2}\; \mu{\rm as}.
\end{eqnarray}
Here, $G$ is the gravitational constant,  and $D=D_{\rm l}D_{\rm s}/D_{\rm ls}$, where $D_{\rm l}$ and $D_{\rm s}$ are the observer$-$lens and observer$-$source distance and $D_{\rm ls}$ is the lens--source distance.  All distances are angular-diameter distances. Lensing causes multiple imaging with separations of up to $\approx 2\theta_{\rm E}$. There is no evidence for multiple imaging in  15\,GHz MOJAVE VLBI observations \citep{2009AJ....138.1874L,2016AJ....152...12L} at 23 epochs between 1995 and 2011. The observations had a resolution of about $1\,$mas. By conservatively assuming that image splitting at $0.5$\,mas would be detectable, we place an upper bound on the Einstein radius: $\theta_{\rm E}<250\,\mu$as. Jets in AG 
usually exhibit moving features that propagate away from the core. If the jet is  directed
close to our line of sight the apparent speed can be superluminal \citep{1984RvMP...56..255B}. In AG at
cosmological redshifts this probes lenses with Einstein radii
$\sim 100 (v_\perp /c)(t/1 {\rm yr})\,\mu$as.

A lower bound on $\theta_{\rm E}$ can be placed using constraints on the size of synchrotron sources. For lensing to yield considerable magnification, the source must be significantly smaller than $\theta_{\rm E}$. Fits to the precise profile of the observed SAV require
\begin{equation}
p=\sigma_s/\theta_{\rm E}\lesssim 0.03.
\end{equation}
However radio sources of a given flux density cannot have a size smaller than a threshold below which inverse Compton losses rapidly ``cool' the relativistic plasma. This sets an upper limit of $\sim 10^{12}$\,K on the intrinsic brightness of an incoherent synchrotron source, though brightness temperatures of $T_{\rm b}\gtrsim 10^{13}$\,K have been observed in compact components in AG jets \citep{2016ApJ...820L..10J}, likely caused by relativistic Doppler boosting. If the source has a Gaussian brightness profile, the brightness temperature can be related to the observed source flux-density by the Rayleigh$-$Jeans law:
\begin{equation}
 S_0 = \mu \frac{ {2kT_{\rm b}}{\lambda^2}}    2\pi \sigma_s^2.
\end{equation}
where $k$ is Boltzmann's constant, $\lambda$ is the wavelength, and $\mu$ is the lensing magnification. Since the Fermat potential near fold caustics has a generic functional form, the peak magnification near a
 fold also has a generic form \citep{1992A&A...260....1S}:
\begin{equation}
\mu_{\rm max} \approx 2.5 p^{-1/2}, 
\end{equation}
where the constant of proportionality has been determined numerically (see Appendix~\ref{sec:model_details}). Equations~1--4 can then be used to put a lower limit on the Einstein radius:
\begin{equation}
 \theta_{\rm E}^2 \ge { { S_o \lambda^2}  \over {2.5 p^{1.5}4 \pi k T_b}} .
\end{equation}
 The mass of the lens can then be determined from $\theta_{\rm E}$ for any given lensing geometry.

\begin{figure*}
\centering
\includegraphics[width=0.8\linewidth]{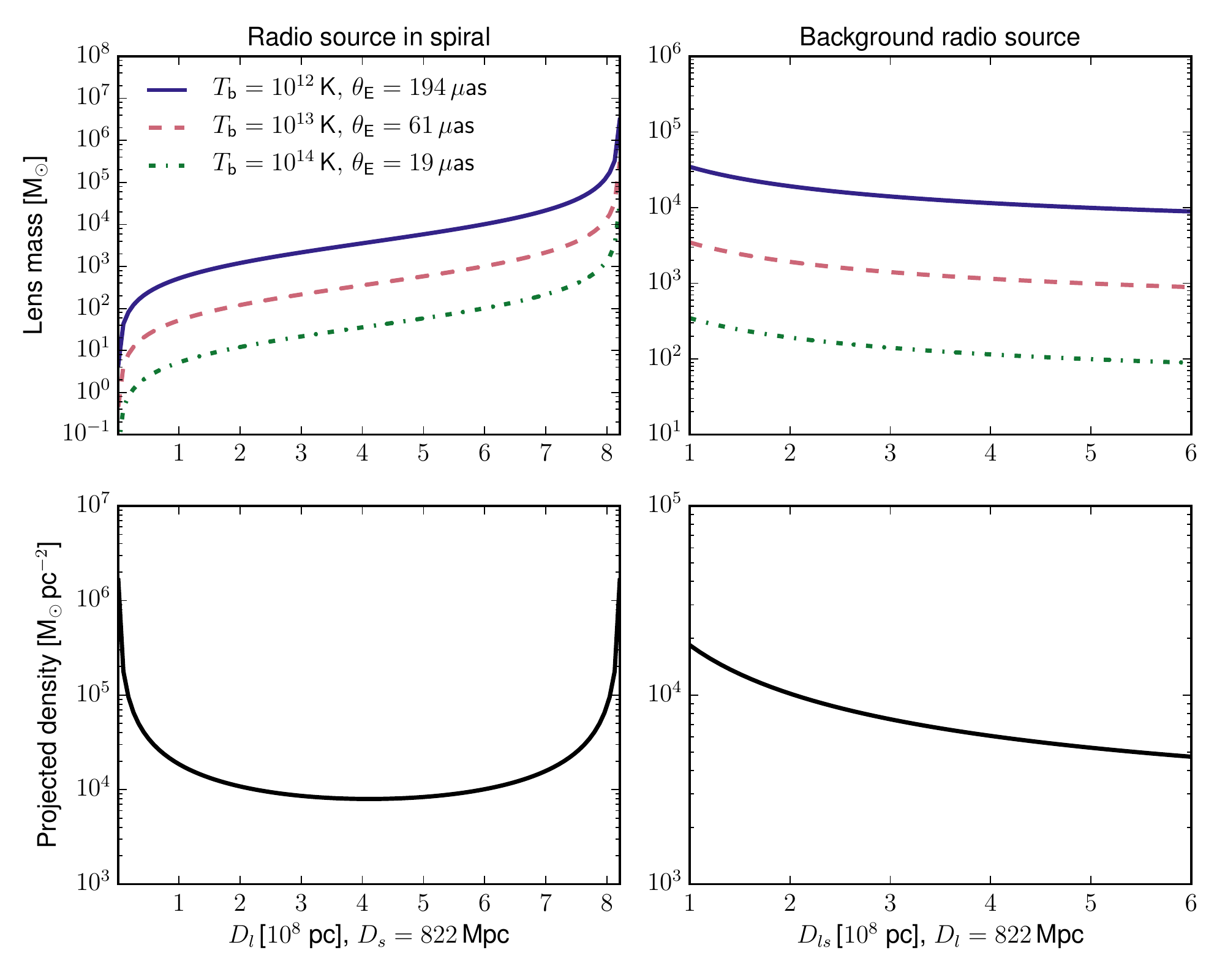}
\caption{Constraints on the minimum lens mass (top panels), and projected mass density (bottom panels). The left and right column correspond to geometries where the radio source is in the putative spiral host, and where the radio source is an unrelated background object respectively. The curves are plotted as a function of observer--lens distance, $D_l$, and the source--lens distance, $D_{ls}$, for the two geometries.\label{fig:m_rho}}
\end{figure*}

In the case of J1415+1320, the peak-to-trough flux-density change for the U-shaped events is $S_o \approx 0.5$ Jy at 15\,GHz and we assume $p=0.03$.  Fig.~\ref{fig:m_rho} shows the lower bounds  of the lens mass and Einstein radius for geometries where the radio  source is located in the spiral and where it is a background radio source.  Since the intrinsic size of the lensed source components are not precisely known, we plot the constraints corresponding to values of $T_{\rm b}$ in the range of $10^{12}$--$10^{14}$\,K. 

\subsection{The location and mass of the lens}
Due to the ambiguity in the location of the radio source (see \S 3), we discuss both possibilities: (i) the radio source is located in the $z=0.247$ spiral, and (ii) the radio source is an unrelated background object.
In the former configuration, we can rule out all models in which the lens lies in our Galaxy: 
Galactic lenses
 consistent with the light-curves of J1415+1320 have a mass range
 consistent with optical micro-lensing surveys.  Such
 surveys \citep{2000ApJ...542..281A} find lensing probabilities of about
 $10^{-7}$, and crossing times of several days, whereas we have found a crossing timescale
 of $\approx 1$ year. If both the source and the lens are in the spiral galaxy, the lens would need to be a $\gtrsim 10^6 M_\odot$ black hole. Since we do not find such objects in the spiral arms in our own Galaxy, we discount this possibility.  
The remaining possibility is that the lens lies between our Galaxy and the spiral host of the radio source.
Here the requirements on the lens mass and projected
 density are $\gtrsim 10^3 M_{\odot}$ and $\gtrsim 10^4 M_{\odot} {\rm pc}^{-2}$.  
As discussed in \S5.1, we obtained a deep
optical spectrum of J1415+1320 with the Keck telescope and
we did not find evidence of an intervening galaxy, although a faint dwarf galaxy could have evaded detection. In the scenario where the radio source is located in the spiral galaxy
at $z=0.247$ there are thus three awkward facts that must be accepted:
(i) J1415+1320 is unique in being a BL Lac object and a CSO located in a spiral rather than in an elliptical galaxy, (ii) there is a 50\,pc offset between the radio source and the optical nucleus, and (iii) the lens is an invisible intergalactic object, possibly a dwarf elliptical galaxy.

\begin{figure}
\centering
\includegraphics[width=\linewidth]{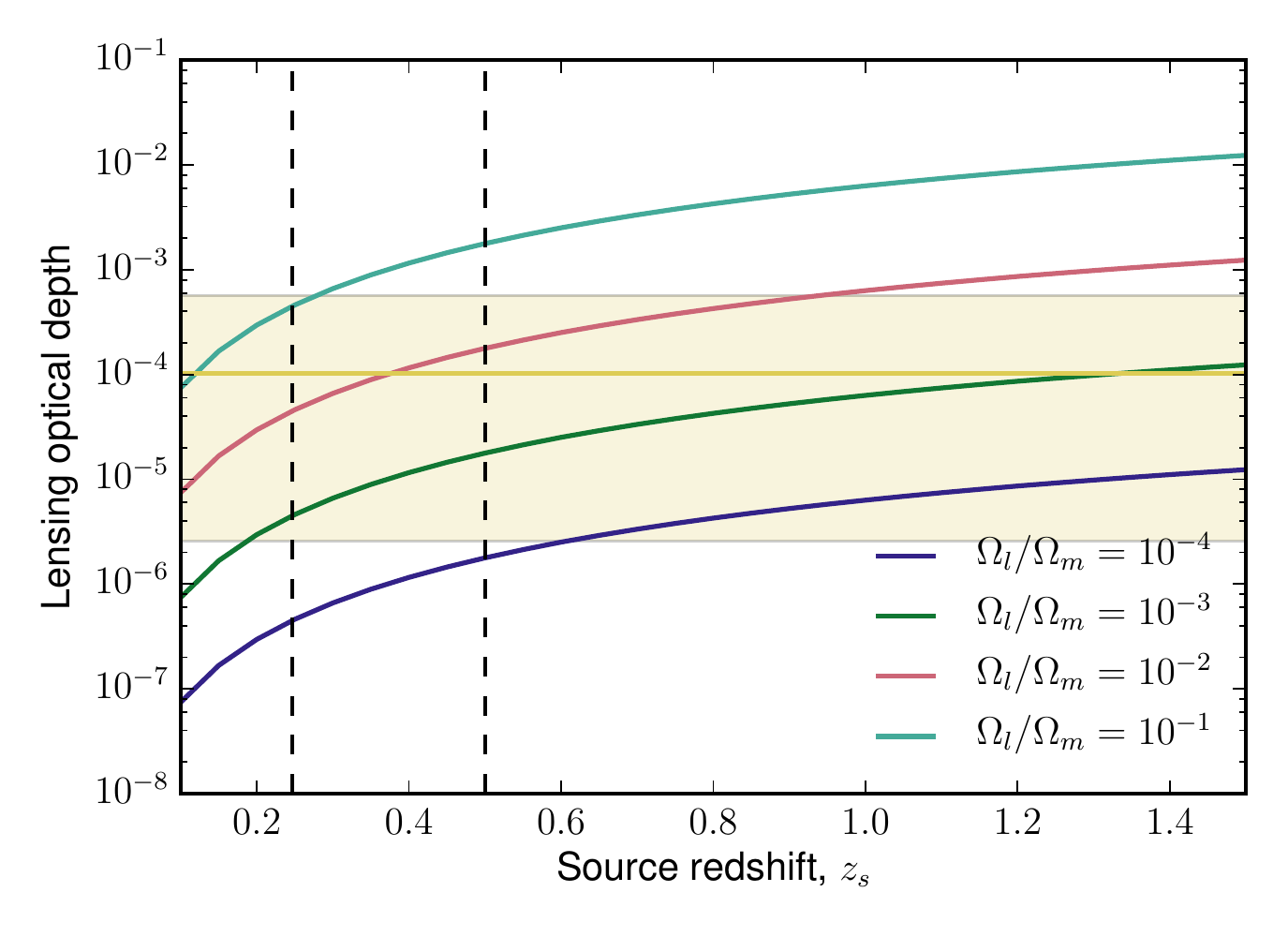}
\caption{Fraction of lines of sight intercepted by the putative lens population (lensing optical depth) as a function of source redshift. The curves correspond to cases where different fractions of  the cosmological matter density are in the lenses. The horizontal yellow line corresponds to the observed probability of intercepting a lens (based on just one sample, J1415+1320), and the shaded region is the corresponding 95\% confidence interval assuming Poisson statistics.\label{fig:tau}}
\end{figure}

In the alternative hypothesis the radio source is a background object and the foreground spiral galaxy provides a natural host for the lens. The absence of multiple imaging on arcsecond scales due to lensing by the spiral's bulge limits the redshift of the radio source to be $z\lesssim 0.5$ \citep{1999MNRAS.309.1085L}. In addition, the angular size of interstellar scattering (ISS) in the spiral must be significantly smaller than the Einstein radius for ISS not to ``quench'' the lensing signatures.  Non-detection of ISS at a wavelength of 18\,cm \citep{1996AJ....111.1839P} shows that any scattering of the  J1415+1320 emission by
the spiral is low compared to several nuclear sight-lines through our own Galaxy at low latitudes.  However ISS in our Galaxy is patchy, and little is known about ISS in other galaxies. In the background source scenario, there are two awkward facts that must be accepted: (i) the low probability of such an alignment, and (ii) the possibly anomalously low ISS, but (i) is an {\it a posteriori} argument and (ii) is based on assumptions regarding ISS that cannot be substantiated by current observations.  For these reasons, the background-source argument is strong, if not compelling, whereas the argument that the radio source is located in the spiral galaxy is weak, but not definitively ruled out.  
In any case, we see from Fig.~\ref{fig:m_rho} that the similar constraints on the lens mass ($\gtrsim 10^3$\,M$_\odot$) and projected density ($\gtrsim 10^4\,$M$_\odot\,$pc$^{-2}$) apply to this geometry too.% 

\section{Discussion}
In this section, we consider the implications of the results obtained thus far and make falsifiable predictions to further test the hypothesis that gravitational lensing is the cause of SAV.
\subsection{Cosmological density of intermediate-mass lenses}
The incidence of lensing may be used to constrain the cosmological fraction of matter in the putative lenses \citep{press1973}. We adopt the {\em Planck} cosmological parameters: $H_0=67.8$\,km\,s$^{-1}$\,Mpc$^{-1}$, $\Omega_{\rm m}=0.308$, $\Omega_{\Lambda} = 1-\Omega_{\rm m}$, $\Omega_{\rm k}=0$. Assuming a flat Universe, this yields a critical density of $\rho_{\rm c} = 3H_0^2/(8\pi G) = 8.633\times10^{-27}$\,kg\,m$^{-3}$.
%where $G$ is the Gravitational constant. 
Let a fraction $f$ of the matter be in the putative lenses of mass $M$. The comoving number density of lenses is then $n=f\Omega_{\rm m}\rho_{\rm c}/M$. Consider a comoving volume element at redshift $z_l$ of ${\rm d}V(z_l) = (c/H_0)(1+z_l)^2 D_l^2/E(z_l)\,{\rm d}\Omega{\rm d}z_l$, where $E(z) = \sqrt{(1+z)^3\Omega_m+\Omega_\Lambda}$. There are, on average, $n{\rm d}V$ lenses in this volume element. Each lens presents a strong-lensing cross-section of $\pi\theta_e^2$. The fractional sky area occupied by all lenses out to the source redshift is therefore
\begin{equation}
\tau(z_s) =  \frac{1}{4\pi}\int n{\rm d}V(z_l)\pi\theta_e^2
\end{equation}
Substituting for $n$, ${\rm d}V$ and $\theta_e$, we get
\begin{eqnarray}
\tau(z_s) &=& \left(\frac{c}{H_0}\right)\int_{z=0}^{z=z_s}\,{\rm d}z\,\frac{Gf\Omega_m}{c^2} \frac{D_{ls}}{D_lD_s} \nonumber \\ &&\times \frac{(1+z)^2}{\sqrt{(1+z)^3\Omega_m+\Omega_\Lambda}}
\end{eqnarray}

Fig.~\ref{fig:tau} shows the expected lensing optical depth for various values of matter densities in the putative lenses $\Omega_l=f\Omega_m$, as a function of source redshift. To estimate  $\Omega_l$, these curves must be compared to the {\em observed} probability of lensing -- one lensed source among a sample of 981 objects, assuming J1415+1320 is unique in the sample. Because lensed objects have larger apparent brightness, they are over-represented in any flux density-limited samples such as the CGRaBS sample. This magnification bias factor largely depends on the faint-end luminosity distribution of the specific source population under consideration, and the properties of the lens population. Both of these are difficult to establish, especially given our postulate of a new population of intermediate mass lenses. We adopt a bias factor of 10 to derive an order-of-magnitude estimate. Accounting for this magnification bias, we find that one in $981\times 10$ sight-lines must be intercepted by the putative lenses. The errors on this lensing rate are large with just one sample, but they can be quantified by assuming Poisson statistics for the incidence of lensing. In Fig.~\ref{fig:tau} we show the 95\% confidence bounds on the lensing rate parameter as a yellow box. Fig.~\ref{fig:tau} shows that for a source redshift of $z_s=0.5$, a 95\% bound of $10^{-4}\Omega_m$ to $3\times 10^{-2}\Omega_m$ may be placed on the comoving density of the putative intermediate mass lenses. Hence, if our lensing hypothesis is true, then a non-trivial fraction of the cosmological matter density must exist in intermediate mass objects with projected densities of $\gtrsim 10^4\,$M$_\odot$\,pc$^{-2}$.

\subsection{Falsifiable predictions}
Our hypothesis that links SAVs to gravitational milli-lensing leads to several key predictions, which we now enumerate: \\
(i) Intensive high-cadence long-term monitoring of J1415+1320 at multiple frequencies should detect further SAV events as new compact source components traverse the lens. The incidence of SAVs should increase with  redshift (see Fig. 8).\\
(ii) During future SAV events, global millimeter-wave VLBI observations should be able to detect creation and annihilation of lensing images. The lensed images form at a typical separation of $\approx 2\theta_{\rm E}\gtrsim 100\,\mu$as which can be resolved by existing VLBI facilities.  Bright image pairs will be closer than $2\theta_{\rm E}$, however they will brighten and merge or fade and separate in a manner that ought to be accessible to millimeter VLBI.\\ 
(iii) The polarization position angle does not change along a ray around caustic crossings, however subimages will be magnified and these can be differently polarized.\\ 
(iv) Finally, if the recently discovered fast radio bursts \citep{2007Sci...318..777L} are at cosmological distances, then they must also be milli-lensed at a rate comparable to AG \citep{2016PhRvL.117i1301M}. In such cases, $\lesssim 10\,$ms time-delays between lensed images of the burst might be detectable with upcoming surveys. 

\subsection{SAV in the BL Lac object AO 0235+164}
\begin{figure*}
\centering
\includegraphics[width=\linewidth]{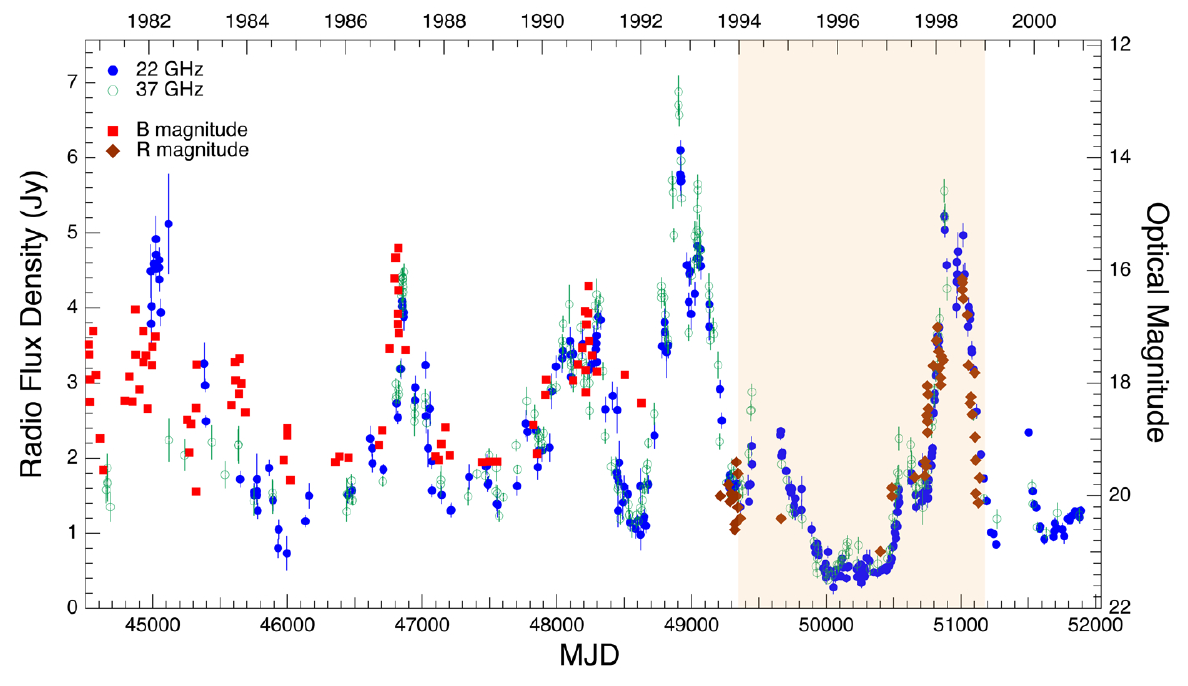}
\caption{The variability in AO 0235+164 at 22 GHz and 37 GHz and B and R magnitudes \citep{2000AJ....120...41W,2009AJ....137.5022N}. The roughly 5-year long putative SAV is highlighted.  \label{fig:AO0235}}
\end{figure*}
The prominence of the two SAV events in J1415+1320 in 2009 and 2014 suggests the possibility that such signatures of radio milli-lensing may have been seen in other objects. For instance SAV may have been seen before, but not identified as such, in the BL Lac object
AO~0235+164. In the  University of Michigan Radio Astronomy Observatory (UMRAO) low-frequency light curves  \cite[Fig.~1]{2003ASPC..300..159A} as well as the 22--90\,GHz light curves \cite[Fig.~2]{2009AJ....137.5022N} there is a characteristic crater U-shaped feature between 1994 and 1999, and the variability was achromatic from radio to optical frequencies \citep{2000AJ....120...41W}, although no optical observations are available for the first half of the crater feature. The 4.8\,GHz, 8\,GHz, and 14.5\,GHz light curves \cite[Fig.~1]{2003ASPC..300..159A} are remarkably symmetric between 1994
and 1999, as are the 22\,GHz, 37\,GHz, and 90\,GHz light curves \citep{2009AJ....137.5022N}. The center of symmetry
shifts to slightly earlier times with increasing frequency. 

In Fig.~\ref{fig:AO0235} we show the Mets\"ahovi 22\,GHz and 37\,GHz light curves and the R and B band light curves \citep{2000AJ....120...41W} at high time resolution.  Here the overall  symmetry between 1994 and 1999 is again clear,
although there are also some departures from symmetry, such as might be caused by intrinsic variability in the radio source flux density and speed, and by the presence of multiple lensed components. The highly correlated variations from radio to optical frequencies over the whole time interval could be produced by one radio component or several, moving behind a caustic network. Gravitational lensing  has been invoked to explain various optical emission features of AO\,0235+164 \citep{1988A&A...198L..13S,2000AJ....120...41W}.  The object has redshift $z = 0.940$ with intervening absorption systems at $z = 0.852$ and $0.524$ \citep{1987ApJ...318..577C}, each of which could have one or several lenses  associated with it.  In fact, given that there are two known intervening systems it would not be surprising to have a complex gravitational lensing system along the line of sight to the radio source that would produce a complex caustic network. Detection of image splitting with high resolution ($\lesssim 100\,\mu$as) global millimeter-wave VLBI observations  is one avenue to ascertaining the presence or absence of milli-lensing in such SAV candidate sources. In addition polarization measurements at radio and optical wavelengths would provide a powerful discriminant between intrinsic variability and gravitational lensing, since the former would likely be chromatic whereas the latter would be achromatic, although the very likely possibility of multiple radio components would complicate the situation.

It is worth emphasizing here that the two objects in which we find the clearest evidence of SAV, J1415+1320 and AO\,0235+164, have both long been suspected of being subject to gravitational lensing \citep{1988A&A...198L..13S, 1992ApJ...400L..13C,1992ApJ...400L..17S, 2000AJ....120...41W}, and that our monitoring observations have led us via the completely new and previously unsuspected phenomenon of SAV to this same hypothesis.

\subsection{Outlook}
We have presented a rare new form of variability in active galaxies that is time-symmetric and achromatic in nature. We propose that such variations are a result of gravitational lensing, wherein compact source components cross the magnification pattern of a gravitational lens at relativistic speeds. Symmetric achromatic variability may therefore provide a powerful new tool for investigating not only the cores of active galaxies but also the putative lenses. These lenses have masses in the range $10^3$--$10^6\,M_\odot$ --- a range that embraces intermediate-mass black holes, cores of globular clusters, dense molecular cloud cores, and compact dark matter halos \citep{1994ARA&A..32..531C}. The incidence of SAV in our sample suggests that a non-trivial fraction of the cosmological mass may be in such lenses, though meaningful constraints cannot be placed with just one or two examples.  

If SAV is confirmed to be due to gravitational milli-lensing it  resolves the longstanding puzzles \citep{1988A&A...198L..13S, 1992ApJ...400L..13C,1992ApJ...400L..17S} over the high brightness and strong variability of the cores of J1415+1320 and AO\,0235+164, as well as the absence of a narrow emission-line region and the low thermal IR flux of J1415+1320, and it would confirm the hypothesis posed some decades ago that some members of the parent BL Lac population are inherently fainter and less variable \citep{1985Natur.318..446O,1990Natur.344...45O}, although these papers were suggesting permanent magnification and not what we are seeing in SAVs.

In the 2014 SAV in J1415+1320 it appears that the millimeter radiation leads the centimeter radiation (Fig.~\ref{fig:light_curve}).  This is not seen in the 2009 SAV, but there the  millimeter sampling is sparse.  High-cadence multi-frequency sampling could confirm or refute this trend in a future SAV in J1415+1320. Similarly, in AO\,0235+164 there are some hints that the optical variations may precede the radio variations (see Fig.~\ref{fig:AO0235}). These hints definitely need confirmation but if confirmed, and if our gravitational lensing hypothesis for SAV behavior is also confirmed, and furthermore if these objects are not anomalous, they could have profound  implications for our interpretation of the relationship between the high-frequency and the low-frequency emission regions observed in AG jets. This behavior is readily interpretable if the moving features are outwardly propagating particle acceleration/magnetic amplification fronts --- shocks if the jet is plasma-dominated and nonlinear hydromagnetic waves if electromagnetic field is dominant. The optically-emitting electrons would cool close to the front and the centroid of their emission would cross any caustic ahead of the radio emission. Early shock-in-jet models \citep{1985ApJ...298..114M} showed that millimeter variations lead cmentimeter variations in AG and would seem to fit this model.  The alternative interpretation of multi-frequency observations of AG jets is that the higher-frequency emission originates further back along the jet towards the core, rather than further out along the jet away from the core, since it is well-known from VlBI observations that this is often the case at centimeter wavelengths in AG \citep{1978Natur.276..768R, 1980IAUS...92..165R}, but whether this is the case in all AG and over a much wider frequency range is by no means clear. The additional resolution that SAV may provide under the lensing hypothesis could well then be the only direct way of exploring this interesting dichotomy other than via sub-millimeter VLBI from space.

The failure to associate dark matter with an elementary particle has revitalized interest in the idea that there may be a large population of primordial  black holes \citep{2016PhRvD..94h3504C}. Further monitoring of relativistically moving radio source components in large samples of radio sources provides a new and sensitive approach to detecting or constraining this putative population.

\acknowledgments
The authors thank Russell Keeney for his tireless efforts in maintaining and operating the OVRO 40 m Telescope.  They thank Sterl Phinney and Shrinivas Kulkarni for useful discussions.  The OVRO 40\,m program has been supported by NASA grants NNG06GG1G, NNX08AW31G, NNX11A043G, and NNX13AQ89G and NSF grants AST-0808050, and AST-1109911. The Submillimeter Array is a joint project between the Smithsonian Astrophysical Observatory and the Academia Sinica Institute of Astronomy and Astrophysics and is funded by the Smithsonian Institution and the Academia Sinica. H. Vedantham is an R.A.\ \& G.B.\ Millikan fellow of experimental physics. T. Hovatta was supported in part by the Academy of Finland project number 267324. R. Reeves gratefully acknowledges support from the Chilean Basal Centro de Excelencia en Astrofisica y Tecnologias Afines (CATA) grant PFB-06/2007.

\facilities{OVRO:40m, Keck:I (LRIS), Mets\"ahovi Radio Observatory, SMA}

\appendix

\section{Lens modeling}
\label{sec:model_details}

\begin{figure*}
\centering
\includegraphics[width=\linewidth]{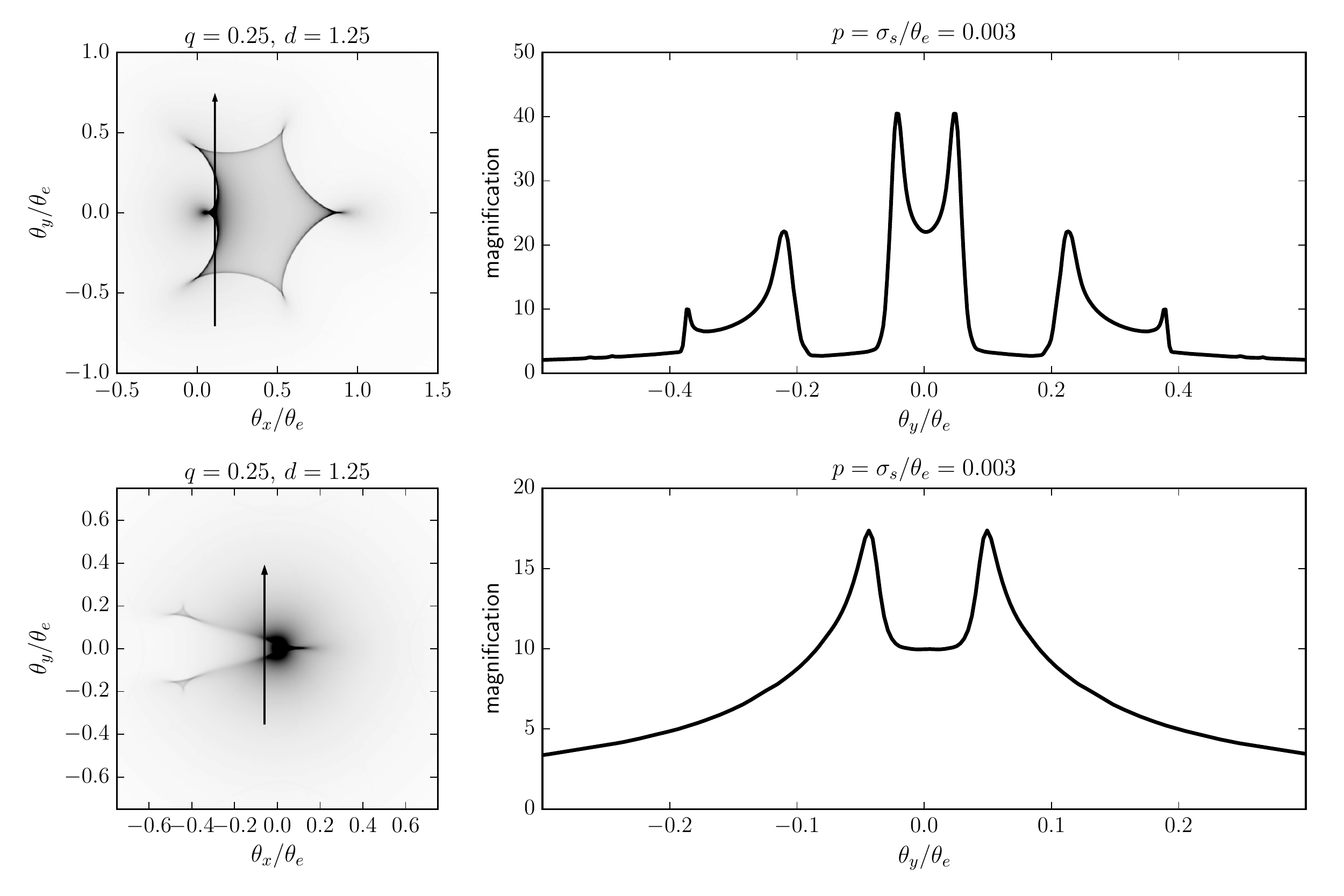}
\caption{Two binary lens configurations (top and bottom rows) that yield the required SRFD$-$FRSD parity. Top-left: Magnification pattern for a binary lens with $q=0.25$, and $d=1.25$ (the ``cupid's bow''). Top-right: Magnification pattern for a source trajectory given by the black arrow. Bottom-left and right: Similar, but for a lens with $q=0.01$, and $d=0.8$ (the `barbed arrow'). The top model can yield the two U-shaped parameters with a single source component, but the viable set of source trajectories is restrictive. The bottom model needs two source components to get two U-shaped events but admits a much larger range of source trajectories.\label{fig:binarylens}}
\end{figure*}

\subsection{Rayshooting simulation}
We employ the usual dimensionless units where all angles are expressed in units of the Einstein radius, $\theta_{\rm E}$. Let the source and lens plane angular co-ordinates (two dimensional) be ${\bf y}=[y_1,\,y_2]$ and ${\bf x}=[x_1,\,x_2]$ respectively. The lens equation relates the image-plane and source-plane co-ordinates of light-rays traveling from the source to the observer. For the single point-mass case, the lens equation is \citet[chap. 8.1.2]{glbook}:
\begin{equation}
\label{eqn:sinlens}
{\bf y} = {\bf x}-\frac{{\bf x}}{\left| {\bf x}\right|^2} \; ,
\end{equation} 
where the second term on the right-hand side is the gradient of the Fermat potential and is the angle by which the light rays are deflected by the lens. Consider a binary mass comprising two point-like objects of mass $1$ and $q$ located respectively at ${\bf 0} = [0,\,0]$ and ${\bf d}=[d,\,0]$. The  lens equation then needs to include the contributions of both masses.
%\begin{equation}
%\label{eqn:binlens}
%{\bf y} = {\bf x} -\frac{{\bf x}}{\left| {\bf x} \right|^2} - q\,\frac{{\bf x}-{\bf d}}{\left| {\bf x}-{\bf d}\right|^2} \; .
%\end{equation} 

The influence of an external smooth mass distribution can be parameterized by the convergence $\kappa$ and shear $\gamma$. The shear in particular accounts for any ellipticity in the external mass distribution.
%i.e. if the mass distribution is more `stretched' along one axis as compared to an orthogonal axis. 
The total deflection is then \citep[chap. 8.2.2]{glbook}:
\begin{equation}
\label{eqn:binlens_cr}
{\bf y} = {\bf x} -\frac{{\bf x}}{\left| {\bf x} \right|^2} - q\,\frac{{\bf x}-{\bf d}}{\left| {\bf x}-{\bf d}\right|^2} - \left( \begin{array}{cc}\kappa+\gamma & 0 \\ 0 & \kappa-\gamma \end{array}\right)\cdot{\bf x}
\end{equation}
where ``$\cdot$'' denotes matrix multiplication. We have chosen the principal axes for the ellipticity of the external mass to lie along the co-ordinate axes, so the off-diagonal elements in the last term of Eq.~\ref{eqn:binlens_cr} are zero. A positive $\gamma$ enhances ray deflection along the $x_1$ axes as compared to the $x_2$ axes; this corresponds to an external mass distribution that is stretched along the $x_2$ axis. Our models therefore make the implicit assumption that the binary axis lies along one of the cardinal axes of the external mass distribution's ellipticity.  Relaxing this assumption extends the parameter space considerably and is not necessary to find adequate models.

\begin{figure*}
\centering
\includegraphics[width=\linewidth]{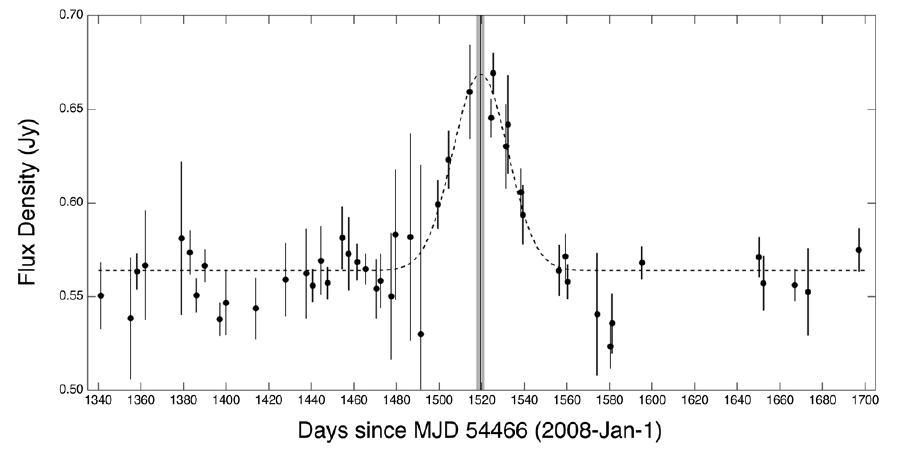}
\caption{OVRO 15\,GHz data for the one-year interval centered on the light curve peak near MJD 54466+1519. The position of the mid-point between the two adjacent peaks (not visible here) and the corresponding uncertainty are shown by the vertical gray bar. This agrees with the centroid of the Gaussian fit to the central peak (dashed line) to remarkable precision ($0.18\%$, see text) and is possibly evidence for an extended caustic network.\label{fig:extended_symm}}
\end{figure*}

\begin{figure*}
\centering
\includegraphics[width=\linewidth]{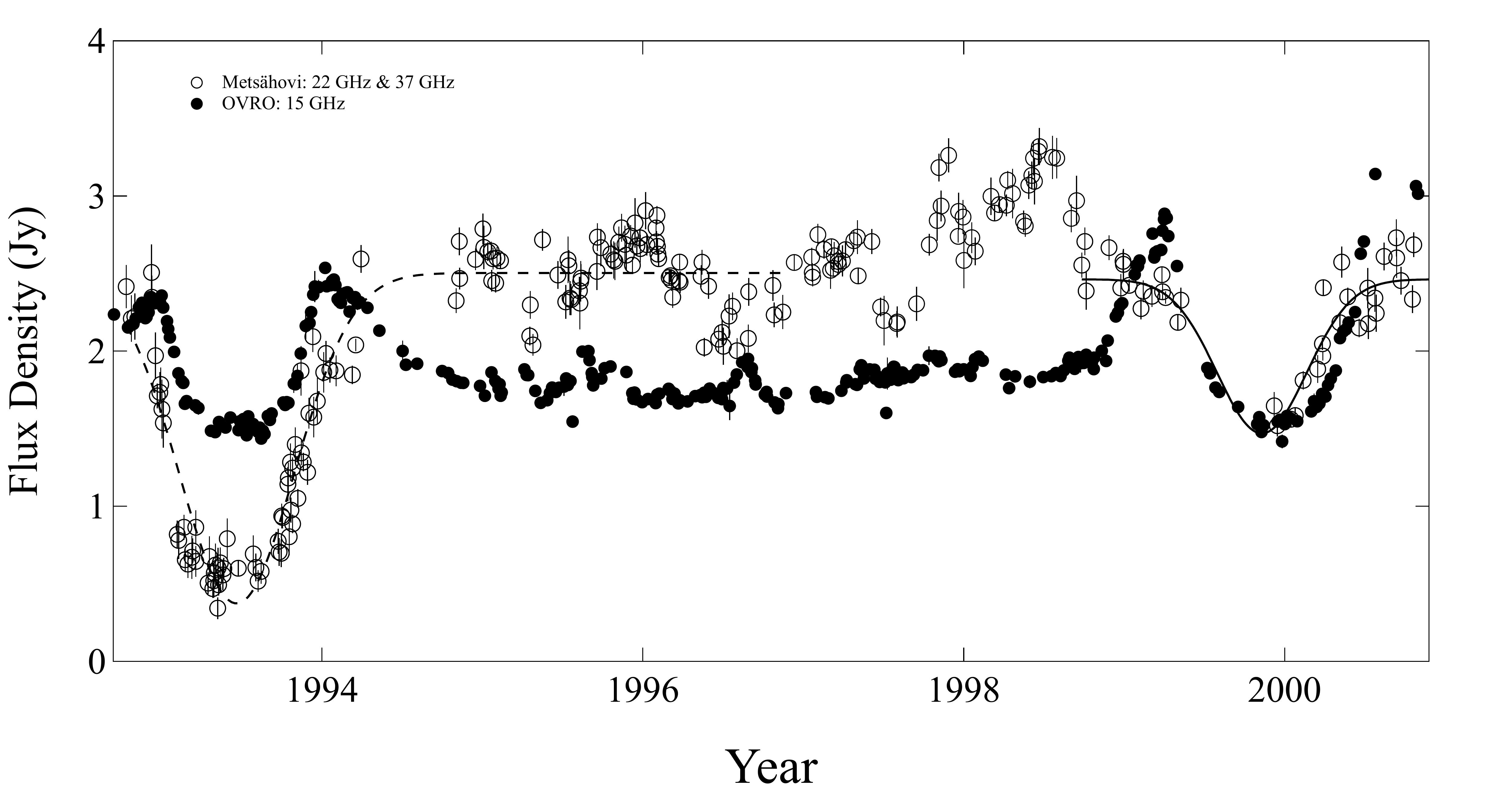}
\caption{Comparison of the two U-shaped features of 1993 and 2000 in J1415+1320 with the two features of 2009 and 2014 after removing flux density trends over these two windows and scaling the time axis of the second window by $\times$1.236.  Large open circles: the Mets\"ahovi 22 GHz and 37 GHz data combined into a single data set with the slopes removed (see text). Small filled circles: the OVRO 15 GHz data with scaled flux density (see text) and with the slope removed and timescale expanded by 23.6\%. Dashed line: Gaussian fit to the 1993 feature at 22 GHz and 37 GHz; Solid line: Gaussian fit to the 2000 feature at 22 GHz and 37 GHz.  These Gaussian fits are not accurate representations of the dip profiles, but illustrate the match between the two data sets. On this model two components ejected 16 years apart trace out a similar pattern as they transit behind the same complex caustic network and differences in the profiles and timescales are ascribed to component flux density variations and a difference in speed of 23.6\%. This figure also illustrates the relative quiescence of the radio source from 2008 to 2016, observed in the OVRO data, which enabled us to make the first identification of SAV. \label{fig:fitfour}}
\end{figure*}

We used equation \ref{eqn:binlens_cr} to simulate the magnification patterns using the technique of inverse-ray-shooting. We follow a large number of rays from the observer to the source plane. We ``shot'' about $10^8$ rays on a regular grid on ${\bf x}$. Using Eq.~\ref{eqn:binlens_cr}, we compute the source-plane location ${\bf y}$ of each ray. The rays are then binned on a suitable source plane grid. Because lensing preserves surface brightness, the magnification is the ratio of the image area to the source area on the sky. The magnification for each source bin is thus the ratio between the number of rays in the bin and the number of rays in the absence of the lens deflection. The lens models used here are simple enough for the computation to be done on a desktop within reasonable compute-time ($<1$\,minute).

\subsection{Rationale for choice of lens configuration}
We endeavored to generate a SRFD--FRSD ``volcano'' type light curve from a lens model that has the following properties: (i) the lenses should resemble commonplace astrophysical objects (point-like masses, elliptical halos, etc.), (ii) the lens-model should be economical -- simple models are favored; and (iii) a rectilinear source trajectory should yield the required SRFD--FRSD symmetry.

We started with the simplest lens of all -- an isolated point-mass lens (axially symmetric gravitational field). Such a lens has a point-like singularity in its source-plane magnification pattern, and therefore yields a FRFD light curve, which cannot adequately describe the data. Elliptical perturbations to a point-like lens typically yield a FRSD--SRFD (crater-type) light curve which is also not viable. Next, we considered binary lenses, which can yield a rich phenomenology of light-curves for various values of the mass-ratio $q\leq 1$ and separation $d$. We computed a library of magnification patterns by varying $q$ and $d$ and identified two scenarios that yield favorable magnification patterns: (i) an asymmetric ($q\ll1$) binary with $d\sim 1$, and (ii) a binary with comparable masses ($q\lesssim 1$) with wider separation $d\gtrsim 1$. The corresponding magnification patterns are shown in Fig.~\ref{fig:binarylens}, along with example source trajectories that give the SRFD--FRSD volcano-type light curve. For brevity, we will refer to the two lens-models as ``cupid's bow'' and  ``barbed arrow'' respectively.

\subsection{External mass sheet}
Having identified two promising candidate lens configurations, we then perturbed the lens parameters to account for (i) the relative timing of the four caustic crossings that together yield the two U-shaped features, and (ii) the flux-density levels throughout the 2008--2016 period. These two constraints can be satisfied by the addition of a smooth background gravitational potential. Such a background potential could reasonably originate in the distributed mass of the galaxy that hosts the lens \citep{chang1979}. 
%It is specified by the convergence $\kappa$ and shear parameter $\gamma$ (more details in \S 2). 
If $\kappa+|\gamma|>0.5$, then the background mass has sufficient lensing strength to independently create multiple images on the macro-scale (typically, arcseconds). Since no arcsecond-level image splitting as been observed in J1415+1320, we imposed the constraint $\kappa+|\gamma|<0.5$. We again simulated a library of magnification patterns with this constraint and found that a large range of ($\kappa,\gamma, q, d$) parameter values can ``fit'' the light-curves. The introduction of external mass is therefore required but is not very constraining. Two examples of favorable lens models that are perturbations of the cupid's bow and barbed arrow lenses are presented in Fig.~\ref{fig:model_fits}. 

\subsection{Model light curve fits to multi-frequency data}
We obtained model light curve fit to the OVRO 15\,GHz data as follows. The data (Fig.~\ref{fig:light_curve}) show a linear decreasing trend in flux density over decade timescales, which may be due to variation in the unlensed source components. We subtracted the linear function of time, $S_{\rm lin}(t) = -1.397\times10^{-4}t+7.922$, from the data, where flux density $S_{\rm lin}$ is in Jy, and time $t$ is MJD in days. Fitting model light curves to the data with the slope removed then requires the following parameters to be estimated: (i) the unlensed flux densities and angular sizes of the lensed components, (ii) the location and orientation of the source trajectory over the magnification pattern, (iii) the Einstein crossing time, $t_{\rm E}$, of the source components, and (iv) the flux density of source components that are not being lensed, assumed to be time-invariant. For simplicity, we assume that (i) all source components move at the same transverse speed along a north--south trajectory though the magnification pattern, and (ii) the source components have a bivariate Gaussian brightness profile with parameter $[\theta^x_s, \theta^y_s]$. We fit the cupid's bow lens with $\theta^x_s=\theta^y_s$, and the barbed arrow lens with $\theta^x_s>\theta^y_x$. Since the source size must be finite, we truncated the source brightness profile at $\pm 2\theta_s$. 

\section{Possible evidence for more extended fold caustic patterns}
\label{sec:extended_caustics}
The basic unit of symmetry we are exploring is the year-long U-shaped feature seen in 2009 and again in 2014 (Fig.~\ref{fig:light_curve}). However, there is possibly another symmetry evident in these data, namely the small peak near MJD 54466+1519, which is very close to the mid-point between the two peaks that straddle the small peak near MJD 54466+1519 in Fig.~\ref{fig:extended_symm}, and are at MJD $54466+742.5\pm 3.9$  and MJD $54466+2295.7\pm1.8$. These locations were determined by fitting low-order polynomials separately to the rising and falling parts of the light curves. The mid-point between these two adjacent peaks is therefore at MJD $54466+1519.1\pm2.1$. We have used the data shown in Fig.~\ref{fig:extended_symm} to determine the position of the central peak by fitting a Gaussian to the central $\sim1$\,yr of data, and we find that the centroid of the small peak lies at MJD $54466+1519.4\pm1.8$. The difference of just $0.3\pm2.8$\, days between these two numbers may be compared with the relatively large interval of $1553$ days between the adjacent two peaks.  The above $1\sigma$ uncertainty amounts to only $0.18\%$ of the interval between the two peaks and may therefore indicate that the two U-shaped features observed in 2008 and 2014 are related by a larger symmetry and that there is in fact a larger caustic network than we have assumed in the main text. Our simple lens models do not offer an explanation for this small peak at MJD 54466+1519, but given the symmetries that proliferate in gravitational lensing we believe that it should  not be ignored at this early stage of exploring the SAV phenomenon. 

Thus it might be the case that the two U-shaped features observed in 1993+2000 and in 2009+2014 are produced by a complex caustic network and that  {\it one} component ejected from the core of J1415+1320 produced {\it both} of the features in 1993 and 2000; and a {\it second} component, %traveling at speed $1.236\times$ slower than the first, 
produced {\it both} of the features in 2009 and 2014.  

If this is the case it is interesting to compare directly the light curve from 1989--2000 with that from 2008--2016.  This comparison is shown in Fig.~\ref{fig:fitfour}.  The mean flux densities at 22 GHz and at 37 GHz from 1989--2000 agree to within $0.2\%$, so we have simply combined these two data sets and removed the slope.  We have also removed the slope from the 15 GHz data in the 2008--2016 window and scaled the flux densities by a factor $\times 2.18$, in order to make the comparison easier to see, and expanded the timescale by a factor $\times 1.236$.  As can be seen in  Fig.~\ref{fig:fitfour}, the two data sets agree quite well so that it is entirely plausible that, allowing for variations in flux density due to intrinsic variability in un-lensed components and variations in lensed components, these patterns could be caused by one component traversing the caustic network in 1993--2000, and a second component
moving at speed $1.236\times$ slower traversing the same caustic network in 2008--2016.  We emphasize here that this interpretation is far from certain.%, and that we present it as an existence proof.
%High resolution ($\lesssim 100\,\mu$as) VLBI imaging may be able to ascertain if this is the case.

\section{Chance alignment probability}
\label{sec:chance_alignment}
Let there be $N_{\rm f}$ foreground galaxies to $N_{\rm s}$ background radio sources and assume both are randomly distributed. We want to compute the expected number of instances where a background source and a foreground galaxy are projected (by chance) to within an angle $\theta_{\rm off}$ from one another. The fraction of sky within a radial distance $\theta_{\rm off}$ from any of the $N_{\rm f}$ foreground galaxies is $\pi\theta_{\rm off}^2N_{\rm f}/(4\pi)$. The expected number of chance-alignments is therefore $N_{\rm f}N_{\rm s}\theta_{\rm off}^2/4$. If however, the population of $N_{\rm s}$  sources are drawn from a flux density-limited sample (such as the OVRO 40-m sample), then those in close alignment with a foreground galaxy will be over-represented due to magnification by the putative milli-lenses. We assume an average milli-lensing magnification of $\mu=10$ (see Fig. 6). For an intrinsic source-count power-law with slope of 1.5 (Euclidean value), the magnification bias factor is $B_\mu = 10^{1.5} \approx 30$. For the OVRO survey, we have $N_{\rm s}=981$, and the required alignment for J1415+1320 is $\theta_{\rm off}=13$\,mas. Based on the fits to the Press$-$Schechter function of \citet{tom2014}, we estimate that there must be about $N_{\rm f}=2\times 10^8$ halos above a mass of $10^9$\,M$_\odot$ up to redshift $z=0.5$ (putative redshift of J1415+1320). Hence the expected number of radio sources in our sample with the required alignment is $B_\mu N_{\rm f}N_{\rm s}\theta_{\rm off}^2/4 \approx 0.006$. While this is  a low value, we remind the reader that this is a case of {\it a posteriori\/} statistics, so the argument against the background source hypothesis is not compelling. If halos of $10^8$ and $10^7$\,M$_\odot$ are also considered as viable foreground systems, then the corresponding number of chance-alignments are 0.03 and 0.2 respectively, although it is unclear if halos in this mass-range will host the putative lenses. 

%\begin{thebibliography}{}
\bibliography{j1415_science_tjp1}
%\end{thebibliography}

\end{document}